\documentclass[12pt,preprintnumbers,superscriptaddress,amsmath,amssymb,nofootinbib]{revtex4}

\usepackage{graphicx}
\usepackage{dcolumn}
\usepackage{bm}
\usepackage{amssymb}
\usepackage{amsmath}
\usepackage{epsfig}    
\usepackage{color}
\usepackage{slashed}
\usepackage[hypertex]{hyperref}
\usepackage{makecell}

\begin{document}

\title{Single and Double SM-like Higgs Boson  Production \\
at Future Electron-Positron Colliders in C2HDMs }

\author{Stefania De Curtis}
\email{decurtis@fi.infn.it}
\affiliation{INFN, Sezione di Firenze, and Department of Physics and Astronomy, University of Florence, Via
G. Sansone 1, 50019 Sesto Fiorentino, Italy}
\author{Stefano Moretti}
\email{S.Moretti@soton.ac.uk}
\affiliation{School of Physics and Astronomy, University of Southampton, Southampton, SO17 1BJ, United Kingdom}
\author{Kei Yagyu}
\email{yagyu@fi.infn.it}
\affiliation{INFN, Sezione di Firenze, and Department of Physics and Astronomy, University of Florence, Via
G. Sansone 1, 50019 Sesto Fiorentino, Italy}
\author{Emine Yildirim}
\email{ey1g13@soton.ac.uk}
\affiliation{School of Physics and Astronomy, University of Southampton, Southampton, SO17 1BJ, United Kingdom}

\begin{abstract}

\noindent {\footnotesize We investigate single- and double-$h$, the discovered Standard Model (SM)-like Higgs boson, production at future $e^+e^-$ colliders 
in Composite 2-Higgs Doublet Models (C2HDMs) and Elementary 2-Higgs Doublet Models (E2HDMs) with a softly-broken $Z_2$ symmetry. 
We first survey their parameter spaces allowed by theoretical bounds from perturbative unitarity 
and vacuum stability as well as by future data at the Large Hadron Collider (LHC) with an integrated luminosity up to 3000 fb$^{-1}$ under 
the assumption that no new Higgs boson is detected. 
We then discuss how different the cross sections can be between the two scenarios 
when $\kappa_V^{}$, the $hVV$ ($V=W^\pm,Z$) coupling normalised to the SM value,  
is taken to be the same value in the both scenarios.
We find that if $\kappa_V^2$ is found to be, e.g., $0.98$, then
the cross sections in C2HDMs with $f$ (the compositeness scale) in the TeV region
can be maximally changed to be about $-15\%$, $-18\%$, $-50\%$ and  $-35\%$ for 
the $e^+e^-\to t \bar t h$, $e^+e^-\to Zhh$, $e^+e^-\to e^+e^-hh$ and $e^+e^-\to t\bar{t}hh$ processes, respectively, with respect to those in E2HDMs. 
Thus, a future electron-positron collider has the potential to discriminate between E2HDMs and C2HDMs, even when only $h$ event rates are measured.
}
\end{abstract}

\maketitle

\section{Introduction}

The discovered Higgs state at the LHC has properties of an isospin doublet field.  
Although many new physics scenarios can be embedded in such a structure, one of the simplest but important examples is a 2HDM which 
naturally includes a neutral Higgs state that can play the role of the discovered one. 
2HDMs are furthermore well motivated theoretically~\cite{branco} and 
generally compliant with past and present collider and other experimental data  while still offering a wealth of new Higgs states.  

However, 2HDMs with elementary Higgses (which we called  E2HDMs) suffer, like  the SM itself, from the so-called hierarchy problem. 
An elegant way to solve it  is to presume that the discovered Higgs boson and its possible 2HDM companions are not elementary 
but rather composite particles to which a pseudo-Nambu-Goldstone Boson (pNGB) nature is assigned. 
C2HDMs embedding pNGBs arising from a new strong dynamics at the TeV scale can be constructed  by explicitly imposing a specific symmetry breaking structure. 
Following Refs.~\cite{DeCurtis:2016scv,DeCurtis:2016tsm}, we will analyse C2HDMs based on 
the spontaneous global symmetry breaking at a scale $f$, typically in the TeV region, of an $SO(6)\to SO(4) \times SO(2)$ symmetry.  
In these C2HDM scenarios there are five physical Higgs states just like in E2HDMs, i.e., 
two CP-even ($h$ and $H$), one CP-odd ($A$) and one pair of charged ($H^\pm$) Higgs bosons. 
As intimated, we identify the (by definition) lightest $h$ state as the discovered Higgs boson with a mass of 125 GeV. 
Within this construct, one can derive  deviations of C2HDM couplings from those of a generic renormalisable E2HDM as well as 
pursue searches for new non-SM-like (composite) Higgs signals different from the elementary case. 
We already considered this aspect at the LHC in Ref.~\cite{DeCurtis:2016tsm}. 

In this paper, we study differences in single- and double-$h$ production cross sections at future $e^+ e^-$ colliders between E2HDMs and C2HDMs, where 
the latter are based on the model construction given in Refs.~\cite{DeCurtis:2016scv,DeCurtis:2016tsm}.
For single-$h$ production, there are three relevant modes: 
(i) Higgs-Strahlung (HS) off a $Z$ boson via $e^+e^- \to Zh$, 
(ii) Vector Boson Fusion (VBF) via $e^+e^- \to e^+e^-h$\footnote{Herein, we neglect considering $W^\pm$ induced VBF as the $hW^+W^-$ coupling scales 
in all scenarios considered in the same way as $hZZ$.} and 
(iii) associated production with  top quarks via $e^+e^- \to t\bar{t}h$. 
The double-$h$ production  can be classified similarly by adding one more Higgs boson $h$ to the final state, 
namely, we have: (i')  $e^+e^- \to Zhh$, (ii') $e^+e^- \to e^+e^-hh$ and (iii') $e^+e^- \to t\bar{t}hh$. 

The single-$h$ production modes are useful to extract the $hZZ$ coupling via (i) and (ii) plus the  $ht\bar{t}$ coupling via (iii). 
Because of the small background cross sections typical of a future $e^+e^-$ machine as compared to those at the LHC, 
one expects to measure these Higgs boson couplings with a good accuracy. 
For example, in Ref.~\cite{ILC-White} the 1$\sigma$ error on the measurement of 
the $hZZ$ and $ht\bar{t}$ couplings at the International Linear Collider (ILC) are expected to be 0.5\% and 2.5\%, respectively, 
for an energy of $\sqrt{s}=500$ GeV and integrated luminosity of $\cal L =$ 500 fb$^{-1}$. 
Notice that, in the $e^+e^-\to t\bar th$ mode, one can also have access to the $AZh$ coupling. 

The double-$h$ production modes are naturally sensitive not only to the $hZZ$ and $ht\bar{t}$ couplings but also 
to the triple Higgs boson coupling $\lambda_{hhh}$. 
In particular, the measurement of $\lambda_{hhh}$ is quite important to reconstruct the shape of the Higgs potential, which has been known 
to be a very challenging task at the LHC~\cite{Gianotti:2002xx,hhh-LHC}. The expected precision achievable
at future $e^+e^-$ colliders in the measurement of $\lambda_{hhh}$ is of ${\cal O}(10\%)$ \cite{ILC-White}. 
This should be contrasted with the much more limited accuracy expected at the LHC, wherein $\lambda_{hhh}$ can be constrained only
within a factor of 3 or so \cite{hhh-LHC-exp} of the SM value.
Furthermore, the heavier CP-even Higgs boson $H$ can contribute to the 
double-Higgs boson production process through its propagators, thereby enabling sensitivity to the $Hhh$ vertex, which is 
expected to be within a factor of 10 or so \cite{Barger:2014qva} in E2HDMs.
In addition, the $AHZ$ coupling becomes accessible alongside the $AhZ$ one
in associated production with top quarks. Finally, quartic couplings of the type $hhZZ$ intervene too. (Notice that  $AhZ$ and $hhZZ$
 are related to the underlying gauge structure and as such are not independent couplings.)    
Therefore, the measurement of the aforementioned double-Higgs boson cross sections at future $e^+e^-$ colliders
is important to also extract information about additional Higgs bosons such as their masses and couplings. 

We will show in this paper that 
there exist measurable deviations induced in C2HDMs  by the dependence upon the compositeness scale in several Higgs couplings which cannot be explained in E2HDMs, no
matter the choice of parameters in either scenarios. 
In particular, assuming a fixed value of $\kappa_V$ 
(defined  by the $hVV$ ($V=W^\pm,Z$) coupling normalised to the SM value), 
the difference between predictions in the two scenarios 
can be even larger than 50\% for the double-Higgs production processes.  
Hence, a future electron-positron machine has the potential to discriminate between E2HDMs and C2HDMs. 

The plan of our paper is as follows.
In the next section, we describe the essential features of C2HDMs that will be dealt with here,
concentrating on the couplings entering the aforementioned production  processes. 
In Sec.~III and Sec.~IV, we tackle the
$e^+e^-$ phenomenology of single-$h$ and double-$h$ production modes, respectively. 
Finally, we conclude in Sec.~V. 

\section{C2HDMs and Their Relevant Interaction Terms}

We give a brief review of our C2HDMs. 
The important parameter which characterises the composite nature of Higgs states is $\xi$, defined by $v_{\text{SM}}^2/f^2$, where 
$v_{\text{SM}}^{}$ is a Vacuum Expectation Value (VEV) related to the Fermi constant via $v_{\text{SM}}^{}=(\sqrt{2}G_F)^{-1/2}\simeq 246$ GeV. 
Therefore, in the $f \to \infty$  limit (or equivalently the $\xi\to 0$ limit) 
all predictions in C2HDMs become the same as those in  E2HDMs. 
In C2HDMs,  $v_{\text{SM}}^{}$ is given by the VEV $v\equiv \sqrt{v_1^2 + v_2^2}$ as 
\begin{align}
v_{\text{SM}}^{} = f\sin \frac{v}{f}. 
\end{align}
The ratio of the two VEVs is expressed by $\tan\beta=v_2/v_1$.

The pNGB Higgses are described by a non-linear $\sigma$-model associated to the coset $SO(6)/SO(4)\times SO(2)$. 
Their effective low-energy Lagrangian ought to be constructed according to the Callan, Coleman, Wess and Zumino prescription, following which 
the scalar potential in  C2HDMs is generated by loop effects. Here, as in Refs.~\cite{DeCurtis:2016scv,DeCurtis:2016tsm},  
we aim at studying the phenomenology of C2HDMs in a rather model independent way by assuming the same general form of this potential as in  E2HDMs 
with a softly-broken discrete $Z_2$ symmetry, 
the latter being imposed in order to avoid Flavor Changing Neutral Currents (FCNCs) at the tree level. For the Yukawa sector, 
we need to assume an embedding scheme for SM fermions into $SO(6)$ multiplets to build the Lagrangian at low energy. 
We adopt here the same setup of ~\cite{DeCurtis:2016tsm}, to which we refer the reader for further details of the model construction.

After an expansion in $\xi$ up to ${\cal O}(\xi)$, we obtain the following interaction Lagrangian relevant to  single- and double-$h$ production:
\begin{align}
{\cal L}_{\text{int}} &= g_{\phi_i VV} \phi_i V_\mu V^\mu + g_{\phi_i\phi_j VV} \phi_i \phi_j V_\mu V^\mu + g_{\phi_i\phi_j V}^\mu \phi_i\phi_j V_\mu 
+y_{\phi_i ff} \phi_i\bar{f}f
+\tilde{y}_{\phi_i ff} \phi_i\bar{f}\gamma_5 f  \notag\\
&+y_{\phi_i\phi_j ff} \phi_i\phi_j\bar{f}f +\lambda_{\phi_i\phi_j\phi_k}\phi_i\phi_j\phi_k, 
\end{align}
where $V_\mu=W_\mu(Z_\mu)$ is a charged(neutral) massive gauge boson, $f$($\bar{f}$) is a SM fermion(anti-fermion) and $\phi_i$ represents a (pseudo)scalar boson. 
Notice that the dimension five term $\phi_i\phi_j\bar{f}f$ appears in C2HDMs because of their non-linear properties.
Using this notation, one obtains 
the (pseudo)scalar boson couplings with the gauge bosons as
 (hereafter, we use the shorthand notations $s_x\equiv\sin x$ and $c_x\equiv\cos x$): 
\begin{align}\label{gauge}
g_{hVV}  &= \left(1-\frac{\xi}{2}\right)c_\theta g_{hVV}^{\text{SM}},\quad g_{HVV} = -\left(1-\frac{\xi}{2}\right)s_\theta g_{hVV}^{\text{SM}}, \\
g_{hhVV} &= \left[1-\frac{\xi}{3}(1+5c_\theta^2) \right]g_{hhVV}^{\text{SM}}, \\
g_{AhZ}^\mu  &= -i\frac{g_Z^{}}{2}s_\theta \left[\left(1-\frac{5}{6}\xi \right)p_h^\mu - \left(1-\frac{\xi}{6} \right)p_A^\mu\right], \\
g_{AHZ}^\mu  &= -i\frac{g_Z^{}}{2}c_\theta \left[\left(1-\frac{5}{6}\xi \right)p_H^\mu - \left(1-\frac{\xi}{6} \right)p_A^\mu\right], 
\end{align}
where $\theta$ is the mixing angle\footnote{In the limit $\xi\to 0$, our $\theta$ is translated in terms of $\beta-\alpha$ as in Ref.~\cite{branco} by 
the replacement: $\theta\to \pi/2-(\beta-\alpha)$ and $H\to -H$. } between $h$ and $H$ and $g_Z^{}=g/\cos\theta_W$. 
For the $AhZ\,(AHZ)$ coupling, $p_\phi^\mu$ ($\phi=h,H,A$) are  the incoming four-momentum. 

The relevant Yukawa couplings are given by: 
\begin{align}\label{Yukawas}
y_{htt}  &=\left[ \left(1-\frac{3\xi}{2}\right)c_\theta +s_\theta\cot\beta \right]y_{htt}^{\text{SM}},\quad
y_{Htt}  =\left[ -\left(1-\frac{3\xi}{2}\right)s_\theta +c_\theta\cot\beta \right]y_{htt}^{\text{SM}},\\
\tilde{y}_{Att}  &= \left(1+\frac{\xi}{2}\right)\cot\beta \, \tilde{y}_{Gtt}^{\text{SM}},\quad
y_{hhtt} = -\frac{2\xi}{3v_{\text{SM}}}\left(2 +\frac{s_{\beta+2\theta}}{s_\beta} \right)y_{htt}^{\text{SM}}. 
\end{align}
In the limit $\xi\to 0$, the coupling $y_{hhtt}$ vanishes: as expected there is no tree level $hht\bar{t}$ coupling in E2HDMs. 
We note that the above expressions of the top Yukawa couplings are common to 
all the four types of Yukawa interactions (I, II, X and Y)~\cite{typeX}, so that there is no type dependence in the cross section of the process where 
only $h$ is mediated such as $e^+e^- \to Zh$ and $e^+e^- \to e^+e^-h$. 
If we consider processes involving the propagators of $H$ and/or $A$, then the type dependence appears in these cross sections through their widths. 
However, we have verified that such dependence is negligible so long that $\tan\beta$ is small (say, below 5 or so), which is a condition we will assume in our analysis.

Finally, the relevant  trilinear Higgs self-couplings are given by: 
\begin{align}
\lambda_{hhh} & = \frac{1}{4v_{\text{SM}}s_{2\beta}}\left[  
(s_{2\beta + 3\theta} -3s_{2\beta+\theta})m_h^2 +
4s_\theta^2 s_{2\beta +\theta}M^2 \right] \notag\\
&+\frac{\xi}{12v_{\text{SM}}}\left[c_\theta m_h^2 + 2s_\theta^2M^2(c_\theta + 2s_\theta \cot 2\beta)\right], \label{hhh}
\\
\lambda_{Hhh} & = \frac{s_\theta}{2v_{\text{SM}}s_{2\beta}}\left[  
-s_{2(\beta+\theta)}(2m_h^2 + m_H^2) + (s_{2\beta}+3s_{2(\beta+\theta)})M^2\right] \notag\\
& + \frac{\xi}{12v_{\text{SM}}}s_\theta \left[m_H^2-2m_h^2 + (1+3c_{2\theta}+6\cot 2\beta s_{2\theta})M^2\right], \label{Hhh}
\end{align}
where $m_H^{}$ and $m_h$ are the mass of $H$ and $h$, respectively, and we fix $m_h=125$ GeV throughout the paper. 
We similarly define the masses of $A$ and $H^\pm$ by $m_A^{}$ and $m_{H^\pm}^{}$, respectively. 
In Eq.~(\ref{hhh}) and (\ref{Hhh}), $M$ describes a soft breaking scale of the $Z_2$ symmetry. 
We note that, in E2HDMs, the $\lambda_{hhh}$ coupling can get ${\cal O}(100\%)$ corrections at the one-loop level without spoiling 
perturbative unitarity as it has been pointed out in Ref.~\cite{hhh-loop}. 
They are due to non-decoupling effects of the extra Higgs boson loops, which become significant when the Higgs masses are mainly 
given by the Higgs VEV, i.e., the $M^2$ parameter is not so large as compared to the (squared) masses of the extra Higgs bosons. 
In our numerical analysis, we do not consider the non-decoupling case. 

\begin{figure}[t]
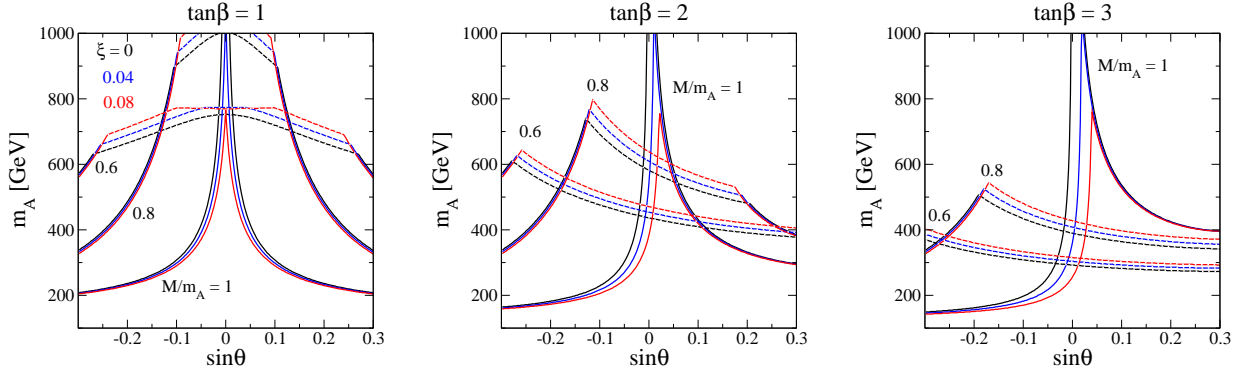

\begin{center}
\includegraphics[width=50mm]{uni_tanb1.eps}\hspace{5mm}
\includegraphics[width=50mm]{uni_tanb2.eps}\hspace{5mm}
\includegraphics[width=50mm]{uni_tanb3.eps}
\caption{Upper limit on $m_A^{}$ ($=m_H^{}=m_{H^\pm}$) from  unitarity  (shown as the dashed part of each curve) and  vacuum stability  
(shown as the solid part of each curve) bounds in the case of $\sqrt{s}=1000$ GeV and $\tan\beta=1$ (left), 2 (center) and 3 (right). 
The value of $\xi$ is taken to be 0 (black), 0.04 (blue) and 0.08 (red). 
We also take the three different values of the ratio $M/m_A^{}$ (1, 0.8 and 0.6) as indicated in the figures. }
\label{uni}
\end{center}
\end{figure}

\begin{figure}[!h]
\begin{center}
\includegraphics[width=0.22\linewidth,angle=270]{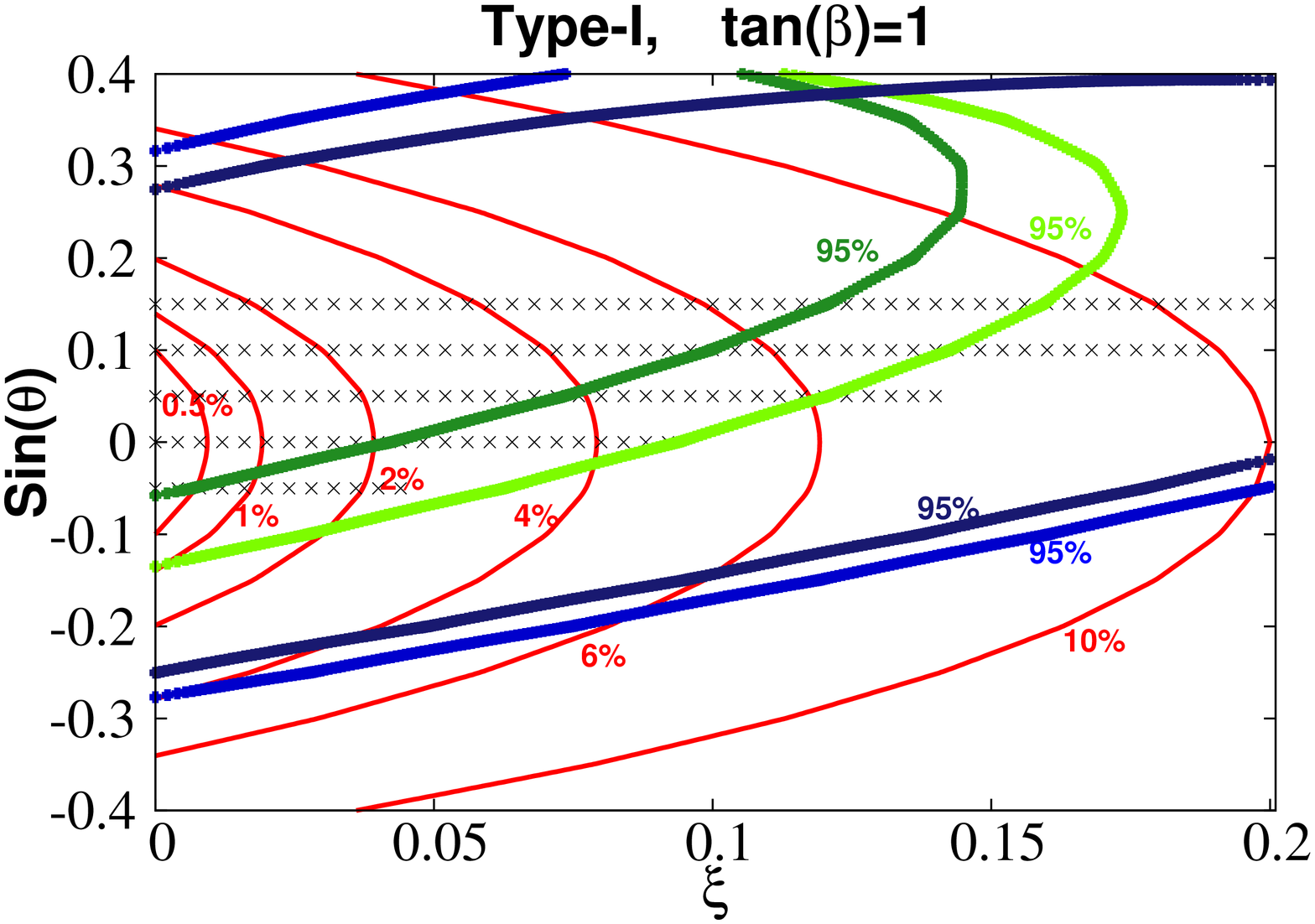}\hspace{2.mm}
\includegraphics[width=0.22\linewidth,angle=270]{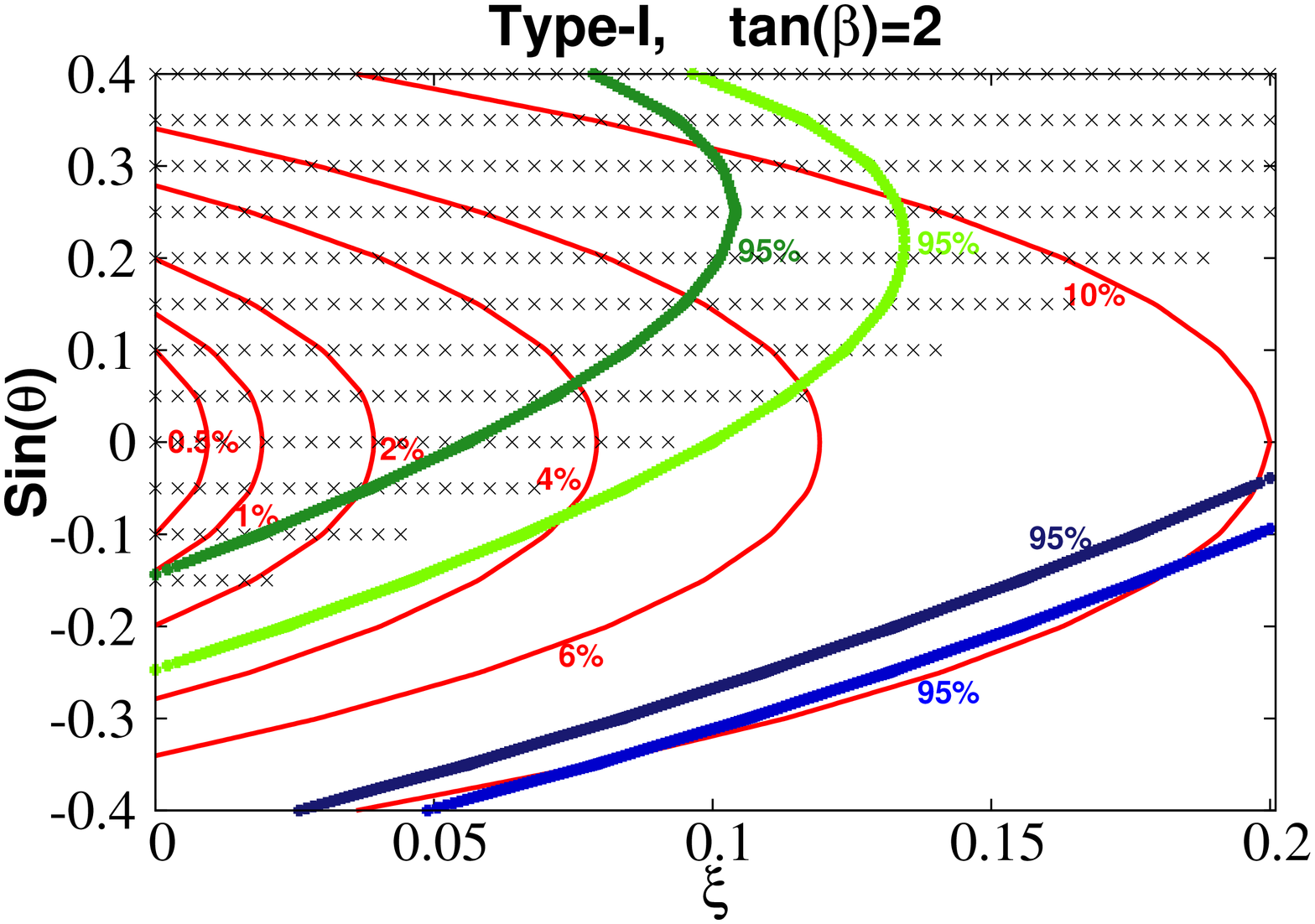}\hspace{2.mm}
\includegraphics[width=0.22\linewidth,angle=270]{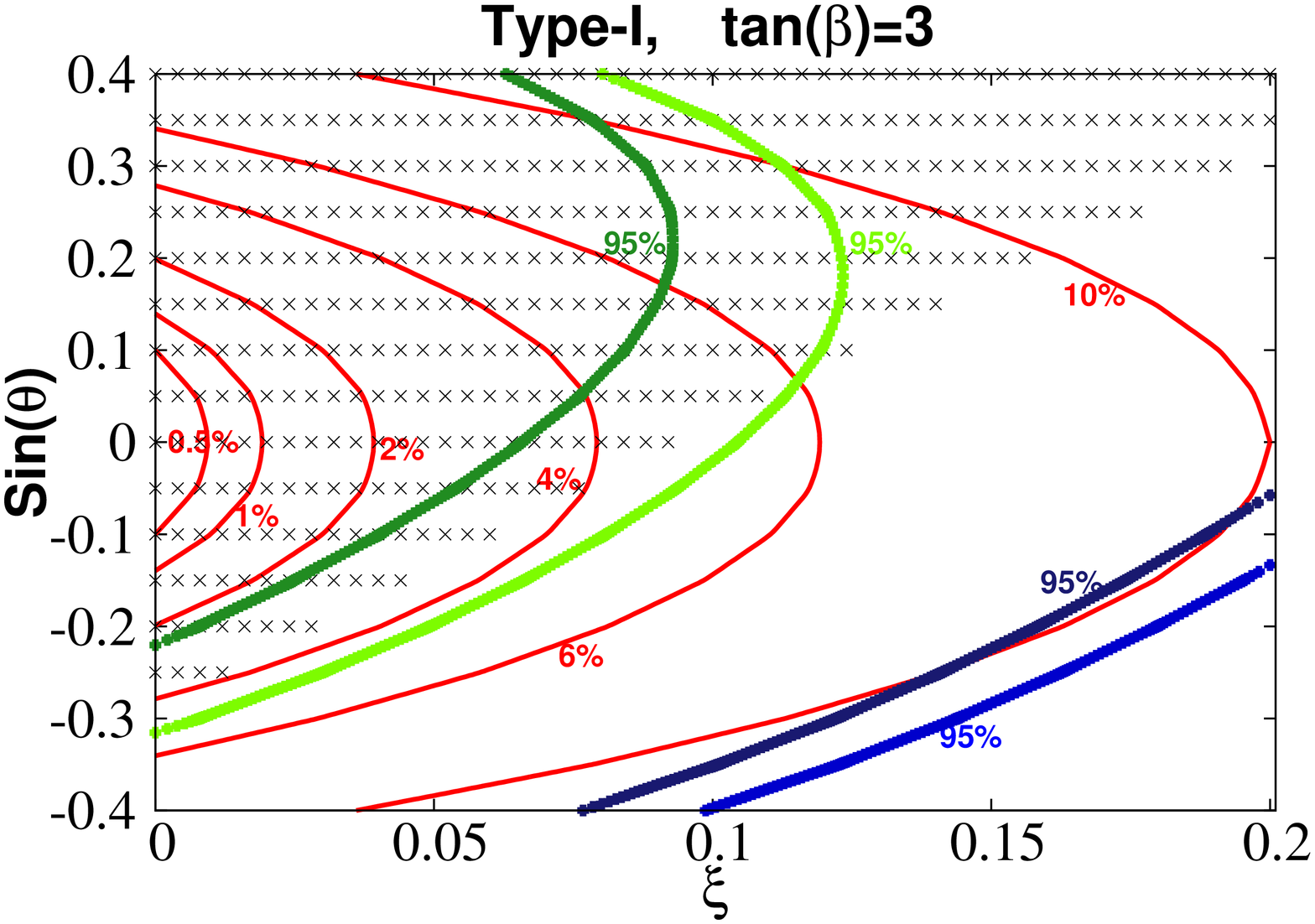}\hspace{2.mm}
\includegraphics[width=0.22\linewidth,angle=270]{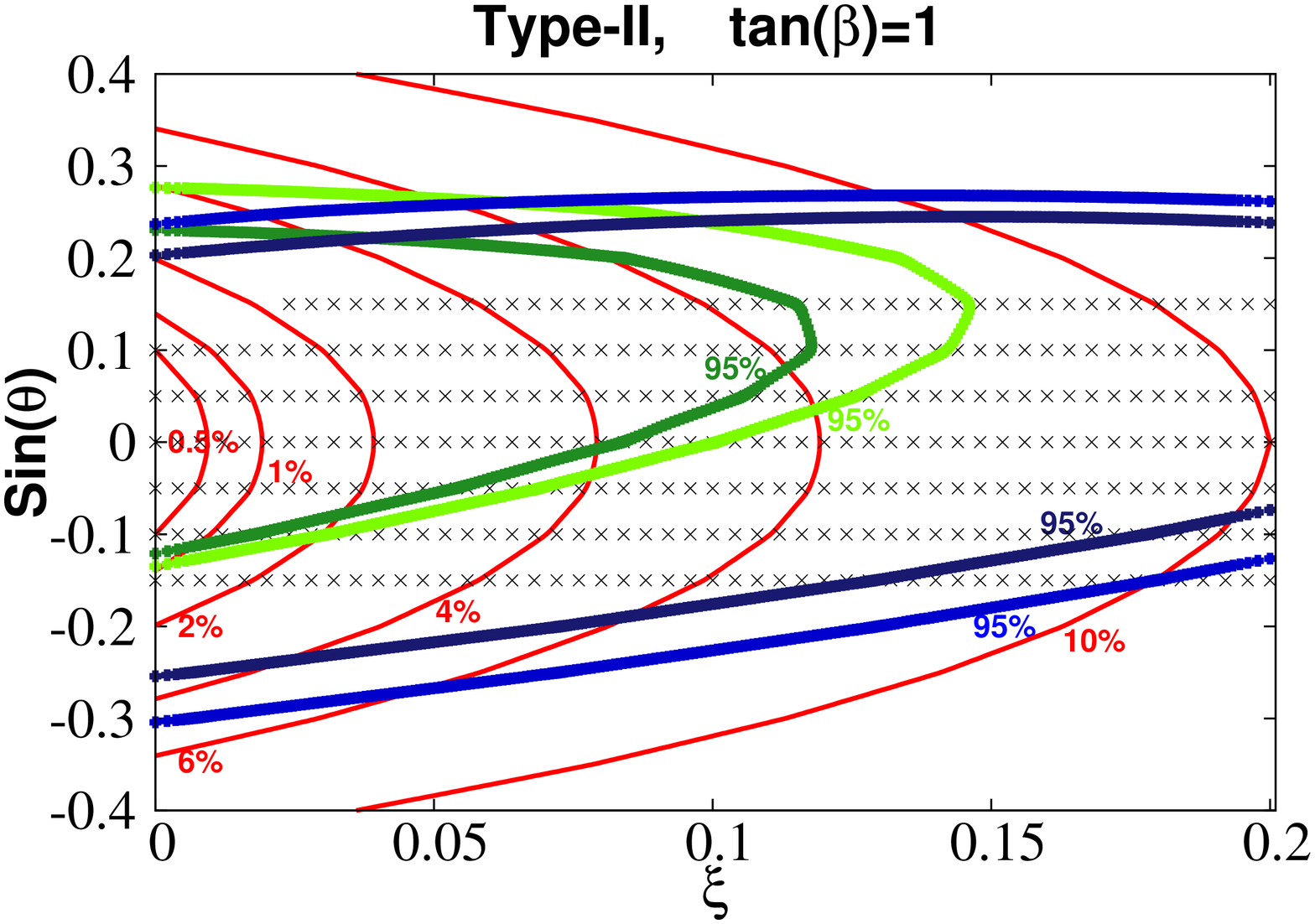}\hspace{2.mm}
\includegraphics[width=0.22\linewidth,angle=270]{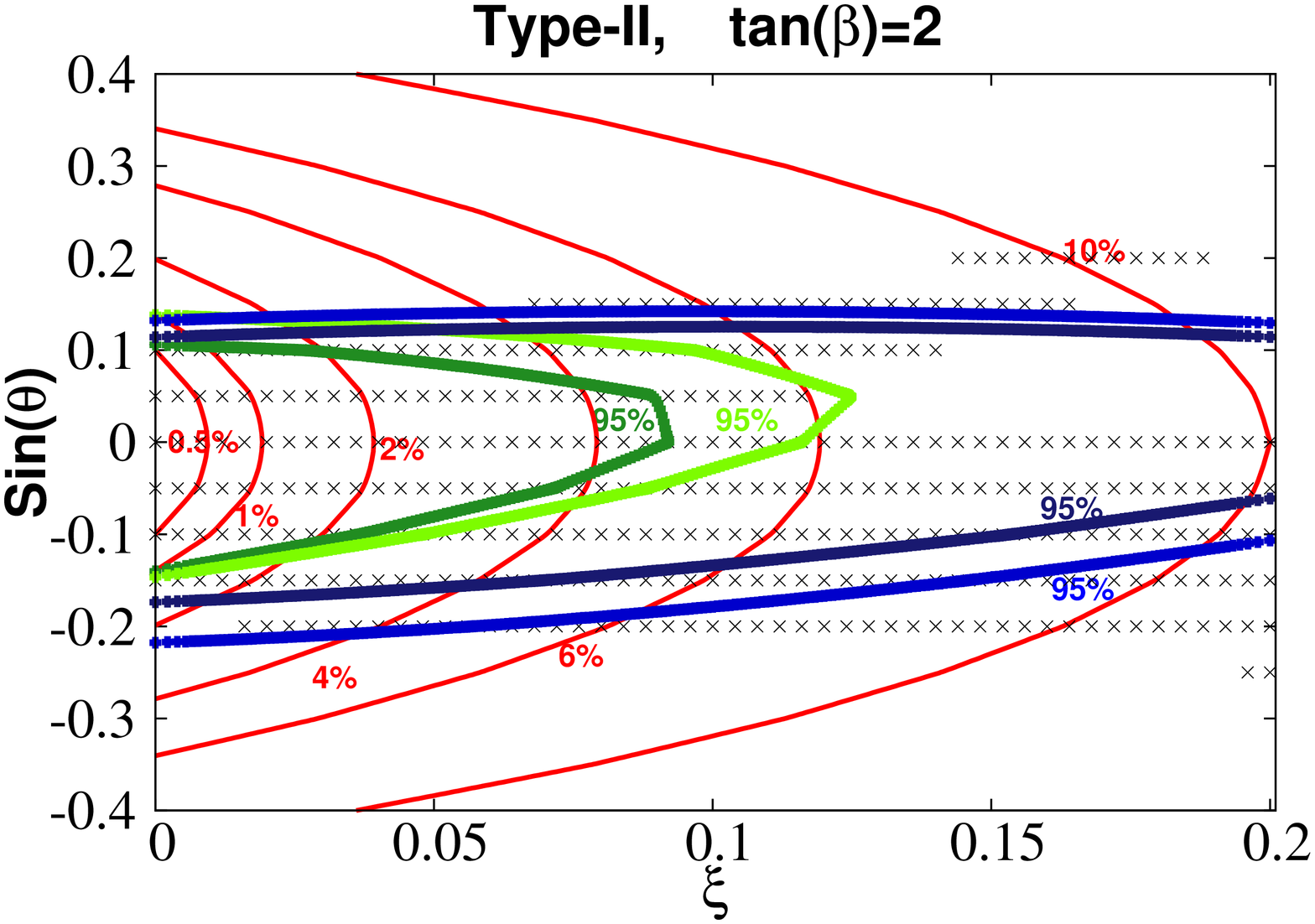}\hspace{2.mm}
\includegraphics[width=0.22\linewidth,angle=270]{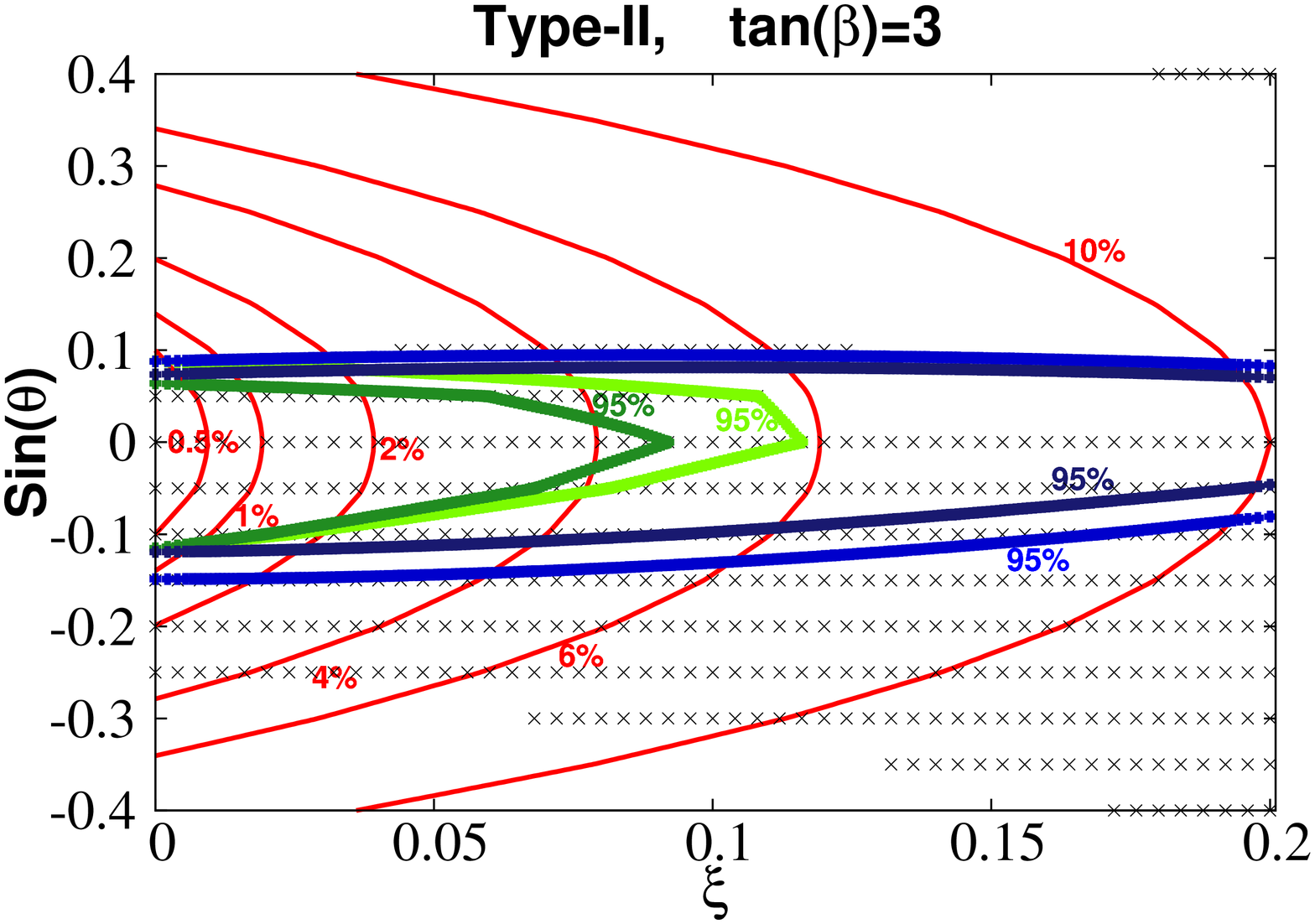}\hspace{2.mm}
\includegraphics[width=0.22\linewidth,angle=270]{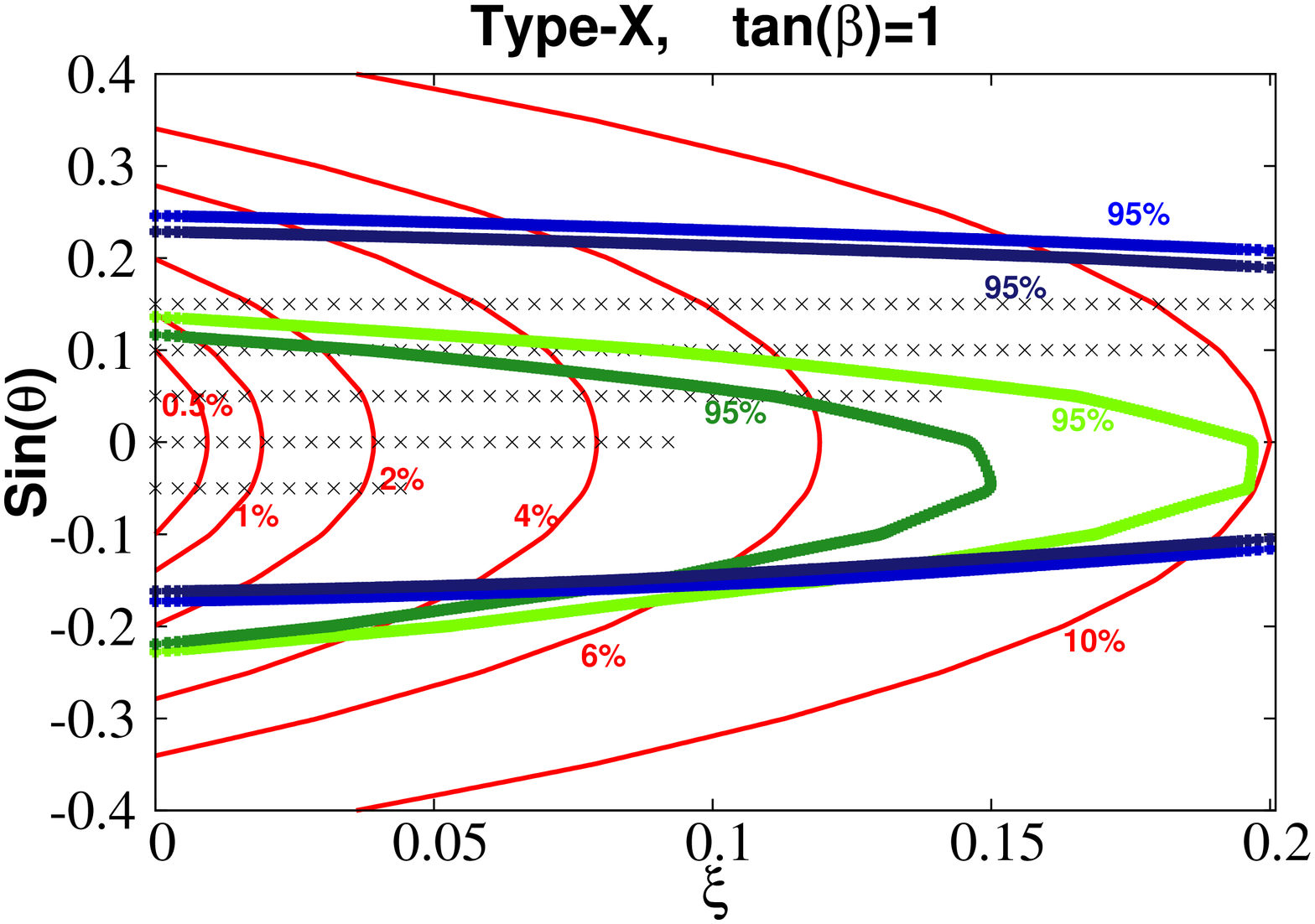}\hspace{2.mm}
\includegraphics[width=0.22\linewidth,angle=270]{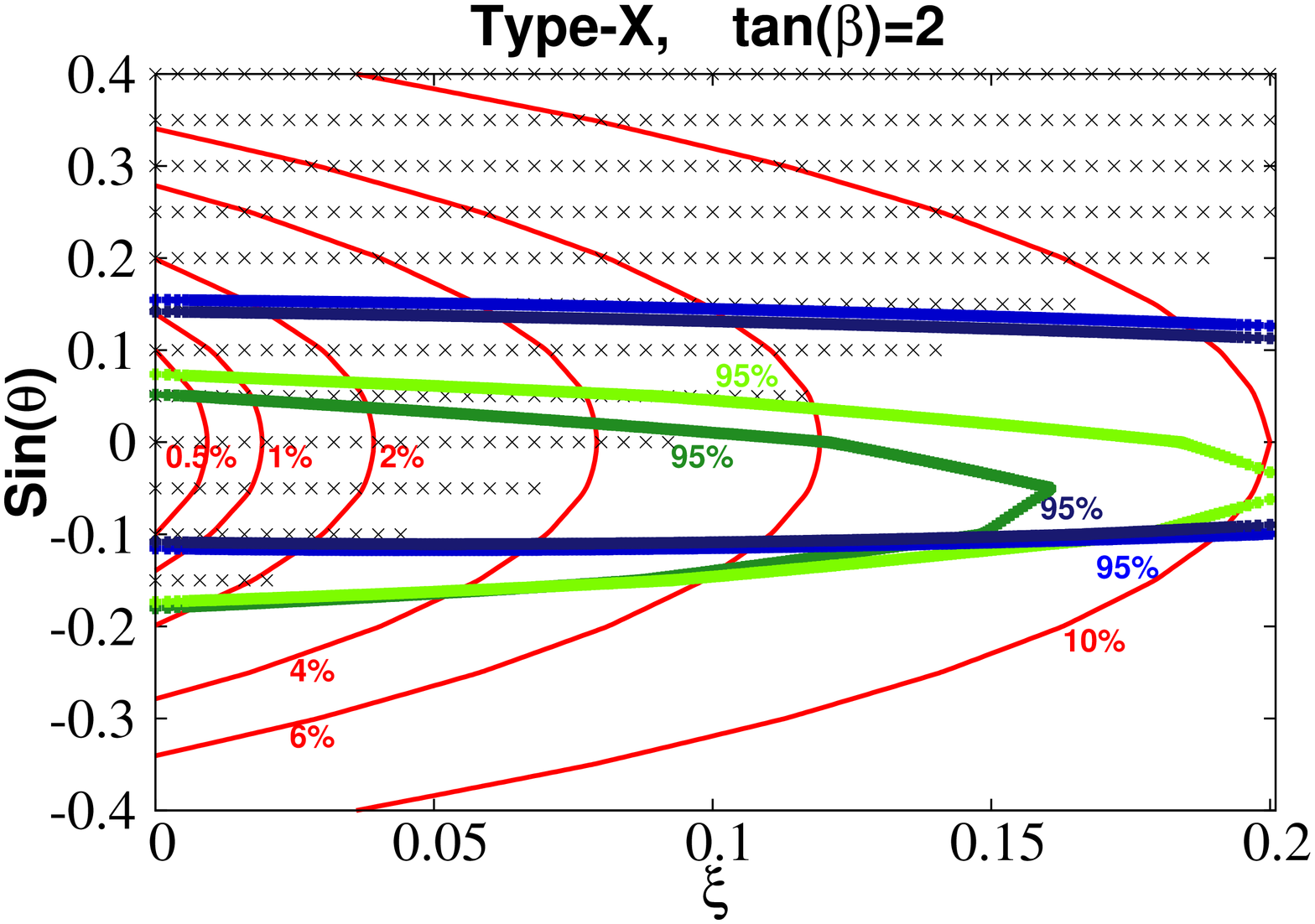}\hspace{2.mm}
\includegraphics[width=0.22\linewidth,angle=270]{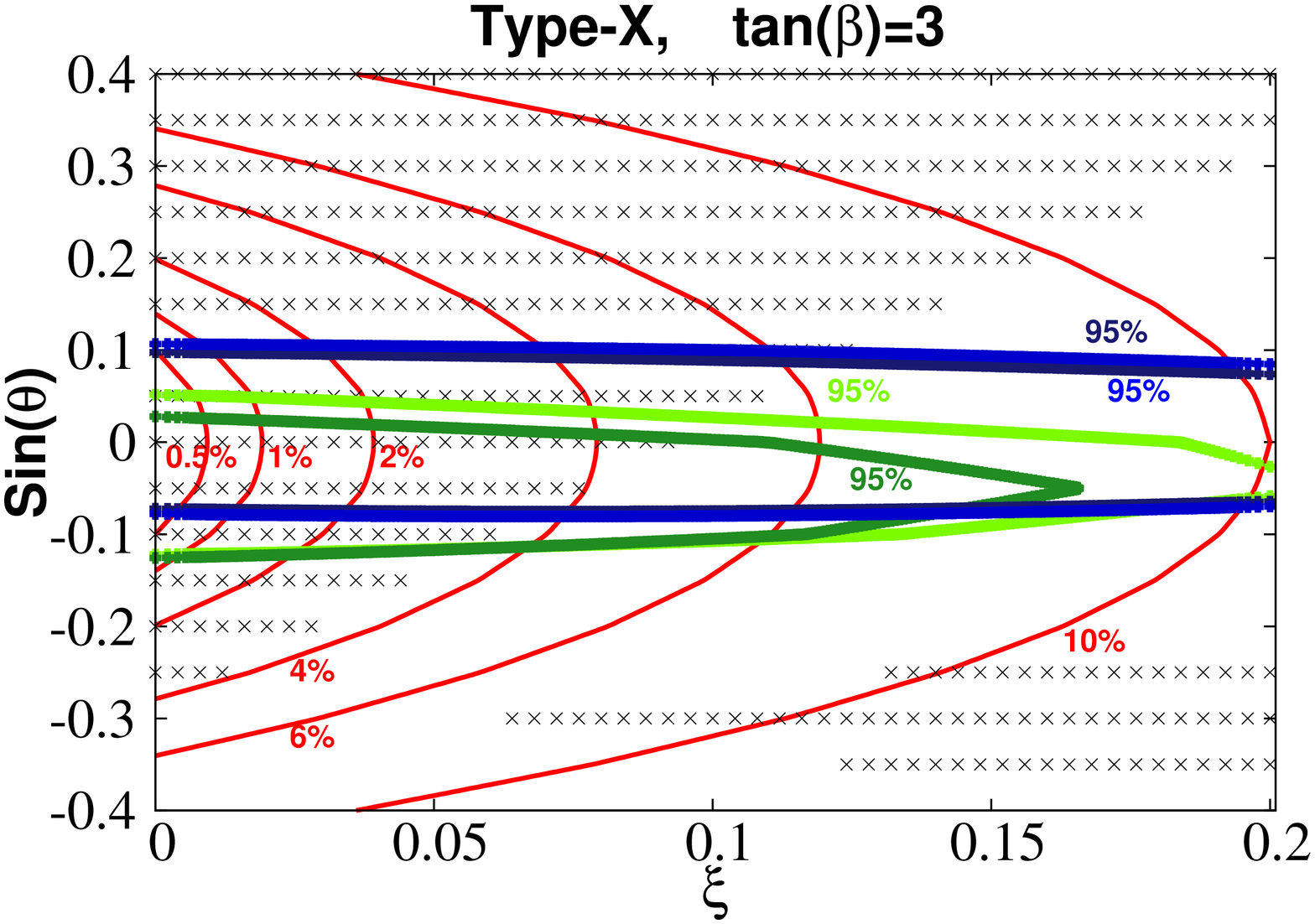}\hspace{2.mm}
\includegraphics[width=0.22\linewidth,angle=270]{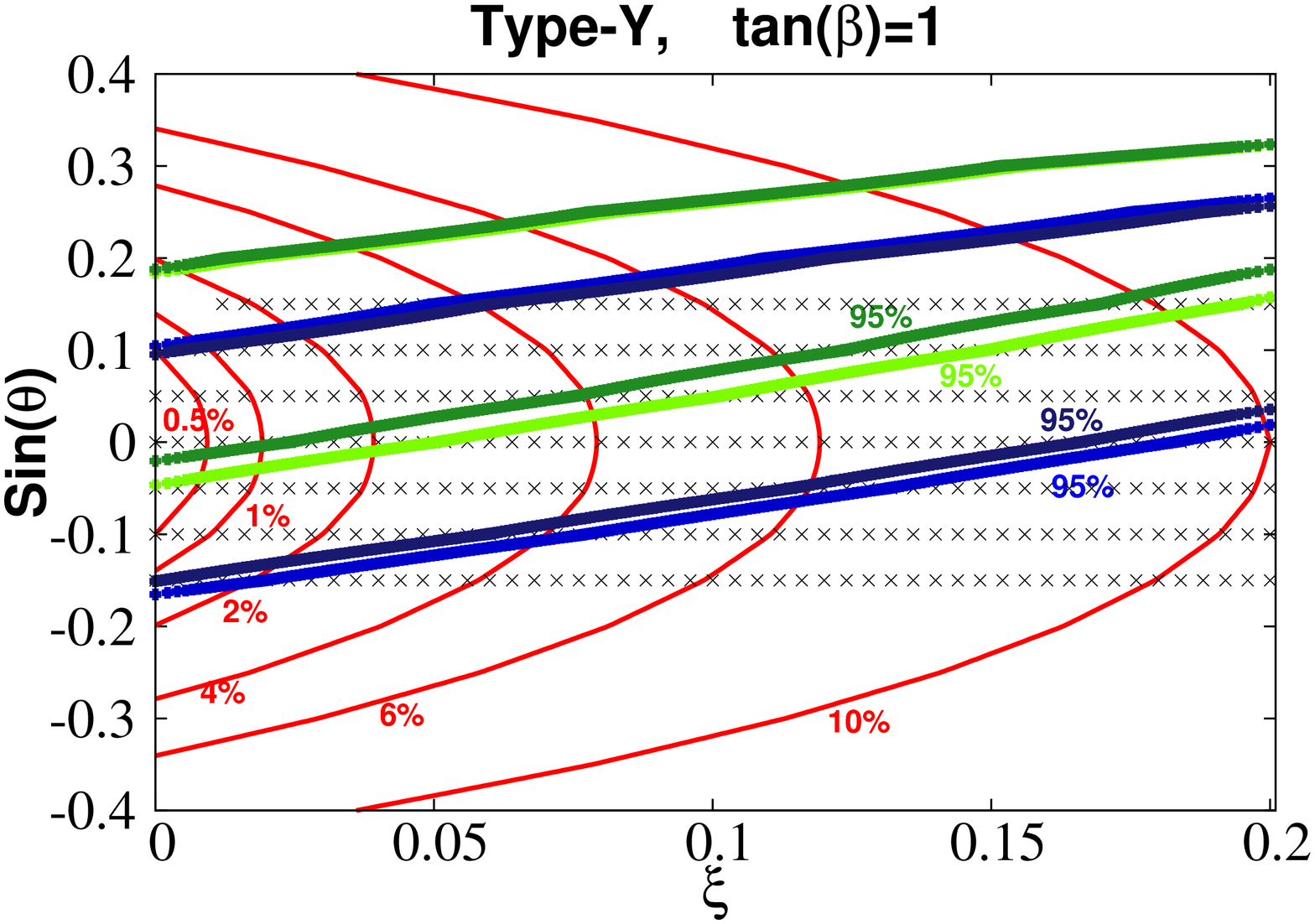}\hspace{2.mm}
\includegraphics[width=0.22\linewidth,angle=270]{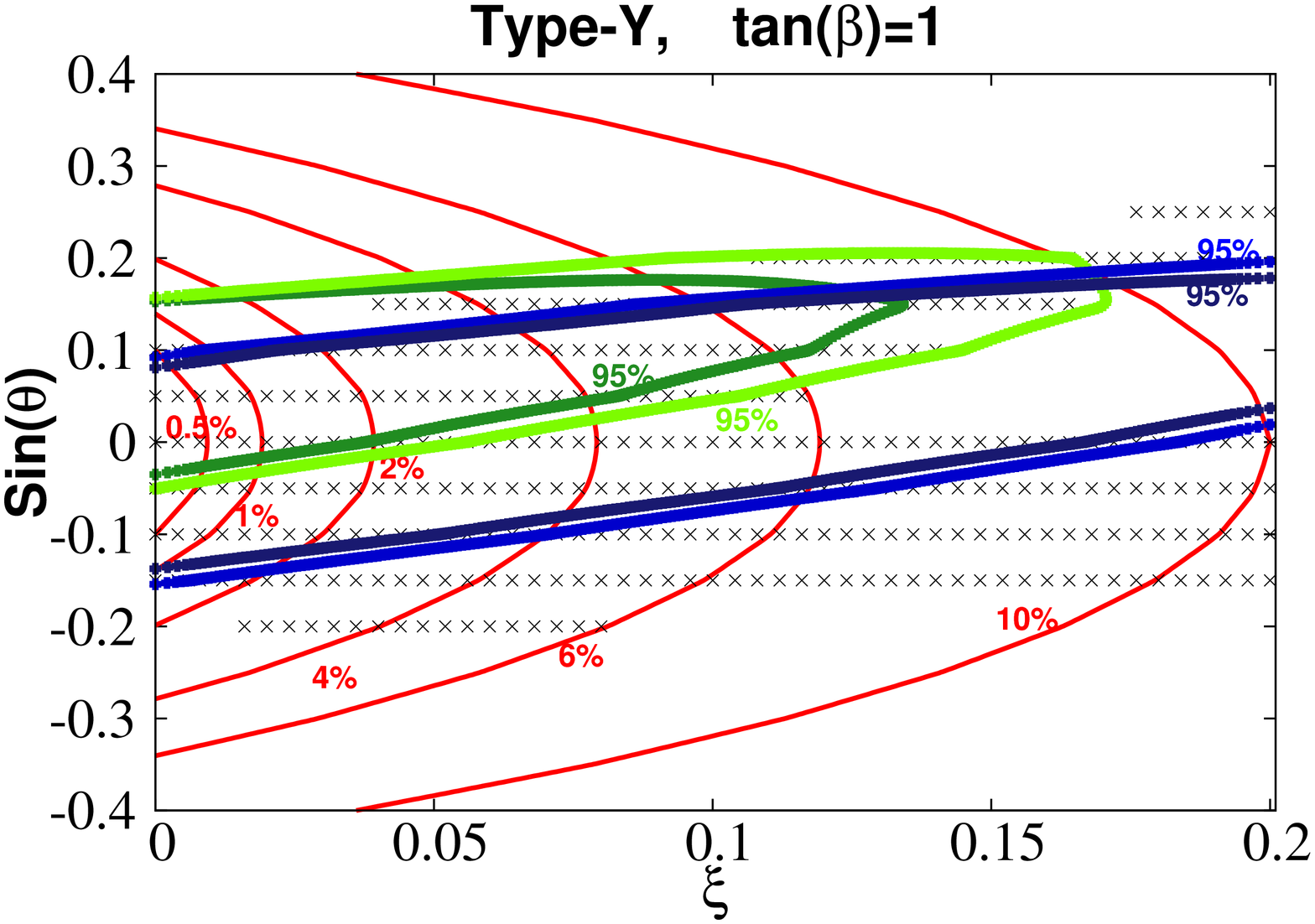}\hspace{2.mm}
\includegraphics[width=0.22\linewidth,angle=270]{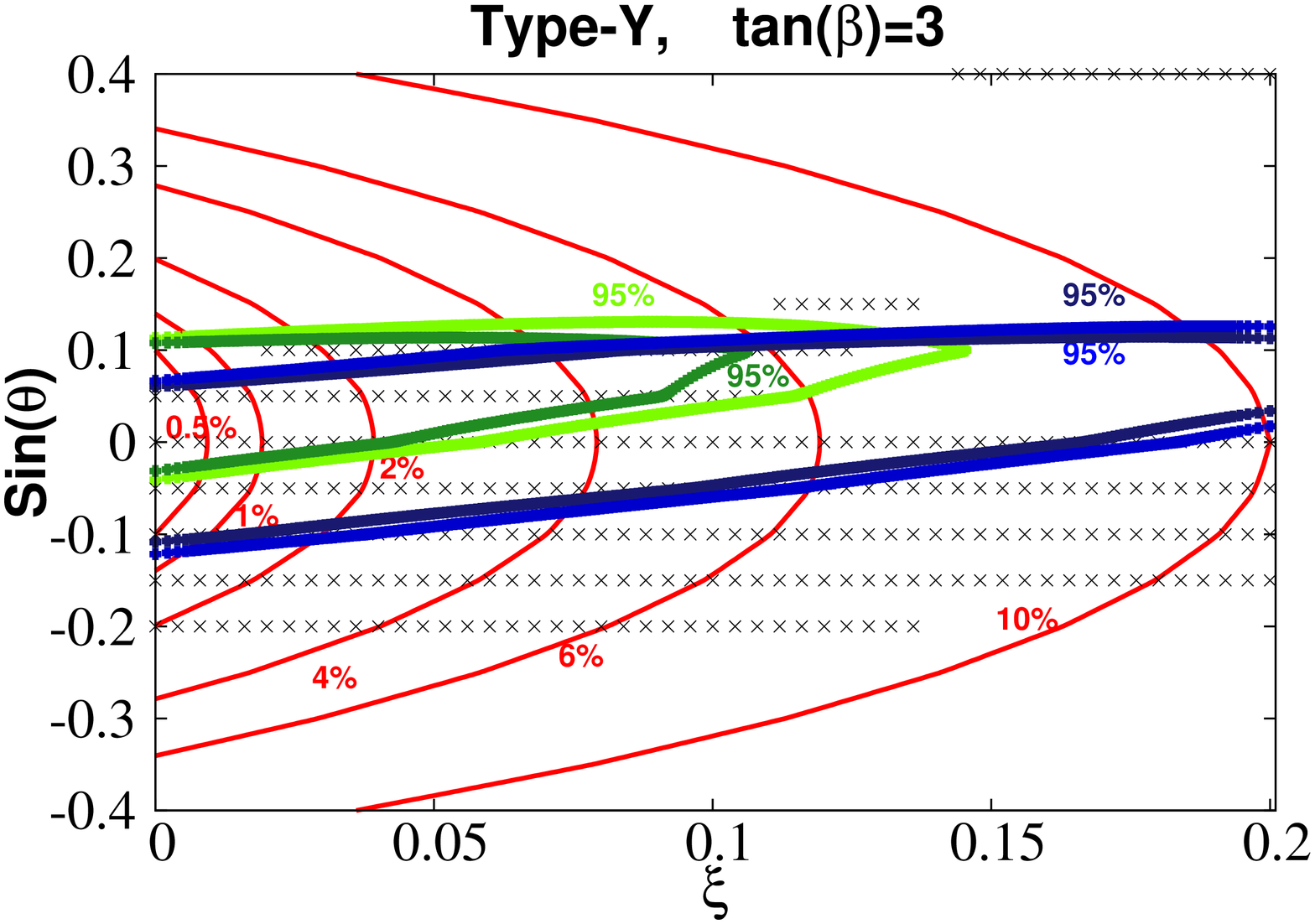}\\ \vspace{5mm}
 \caption{Regions marked by x are allowed by the LHC data at 95\% CL by using the HiggsBounds tool.  
The four rows show the results in the C2HDM of Type-I, Type-II, Type-X and Type-Y. 
The light- and dark-green (light- and dark-blue) curves 
display  the compatibility with observed Higgs signals (SM signal strengths) at $ \Delta\chi^{2}=$6.18 (95.45\% CL)  extrapolated to
  300 fb$ ^{-1} $  and  3000  fb$ ^{-1} $ of luminosity respectively. 
Red curves are contours of $|\Delta \kappa_V|=|g_{hVV}/g_{hVV}^{\rm SM}-1|$. 
The input parameters are chosen to be $m_{H}=m_{A}=m_{H^{\pm}}=500$ GeV and $ M=0.8 m_A^{} $. 
The left, center and right panels indicate  $ \tan\beta $=1, 2 and 3, respectively. }
 \label{HB/HS}
 \end{center}
 \end{figure}

 \begin{figure}[!h]
 \begin{center}
\includegraphics[width=0.22\linewidth,angle=270]{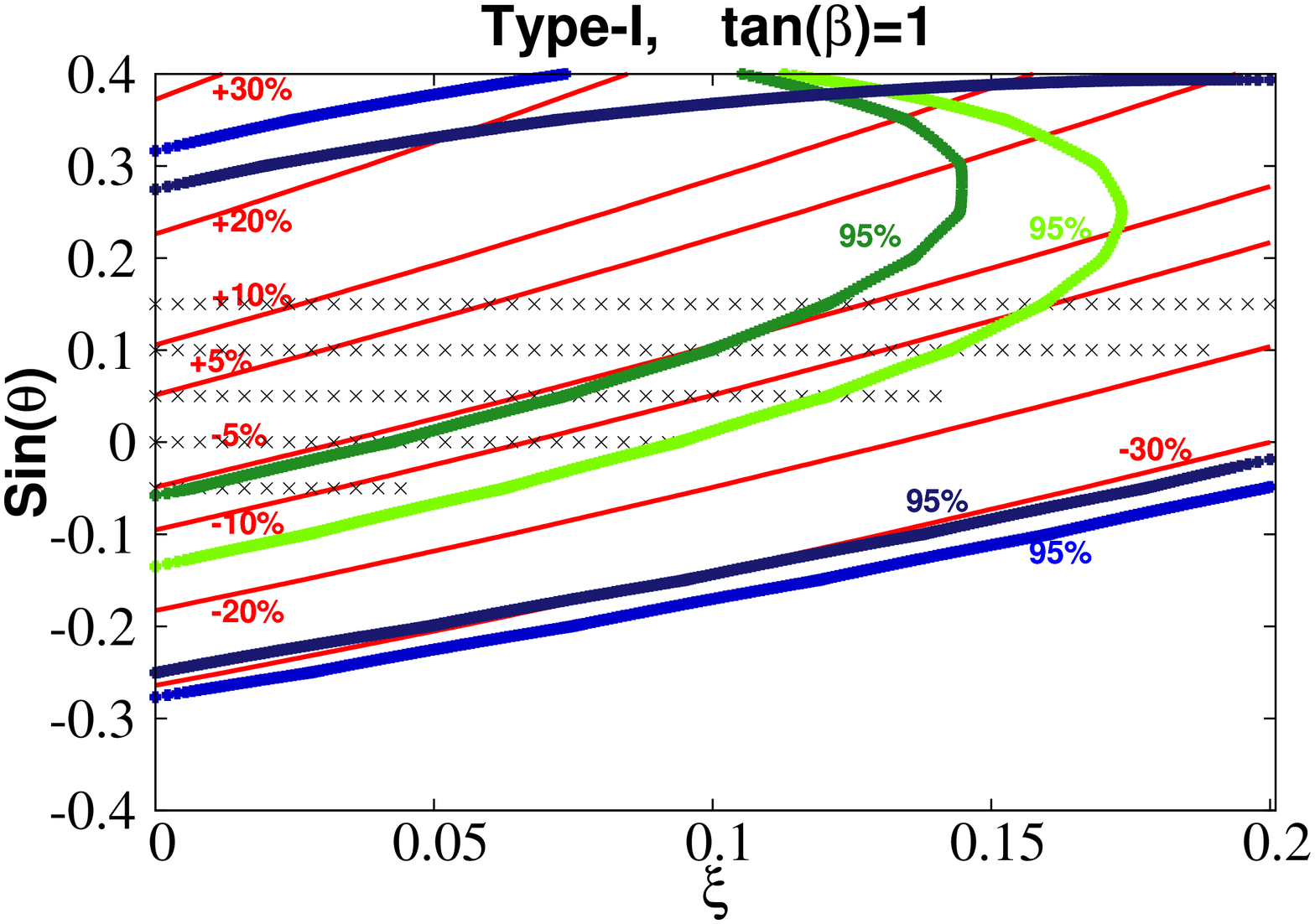}\hspace{2.mm}
\includegraphics[width=0.22\linewidth,angle=270]{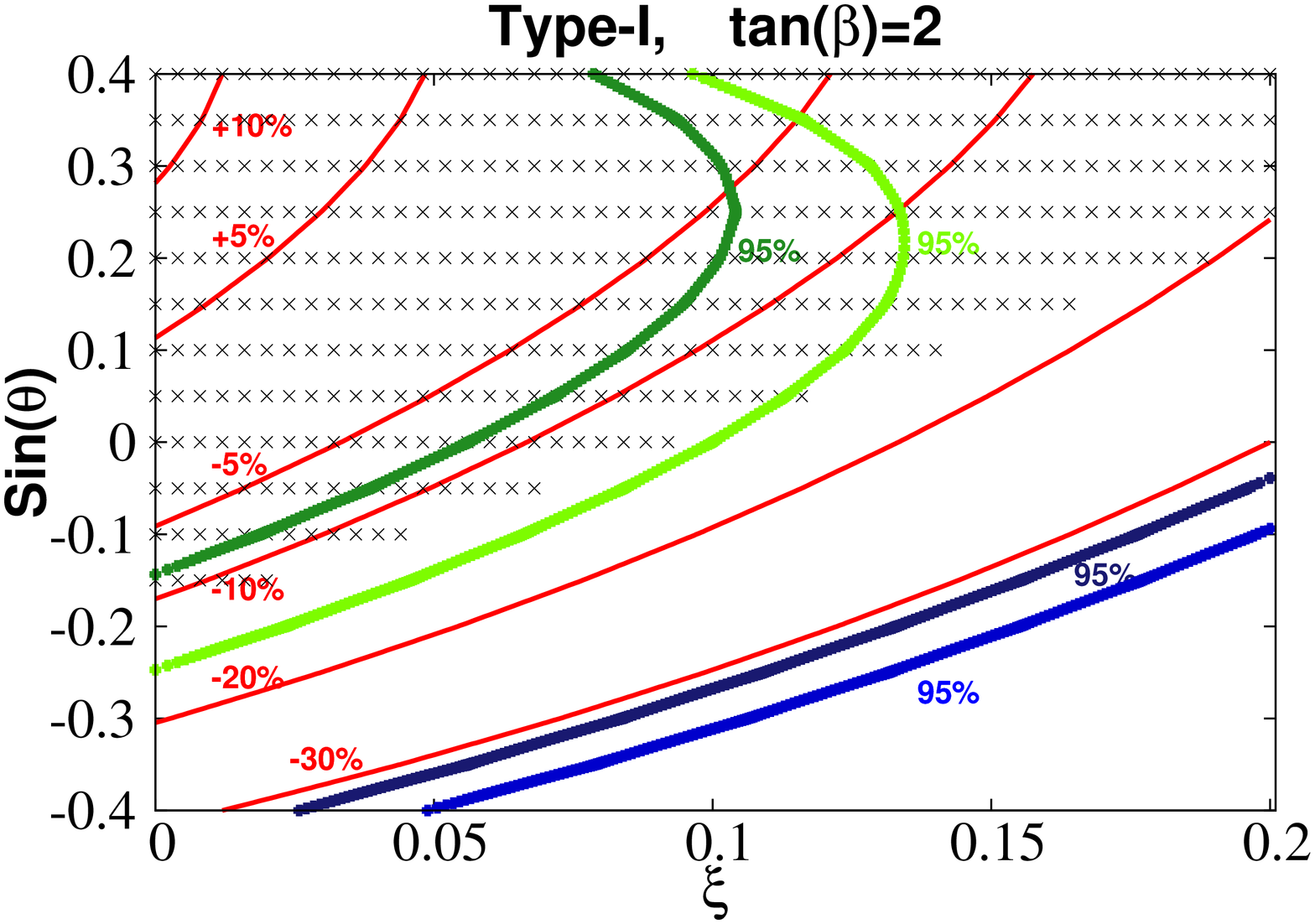}\hspace{2.mm}
\includegraphics[width=0.22\linewidth,angle=270]{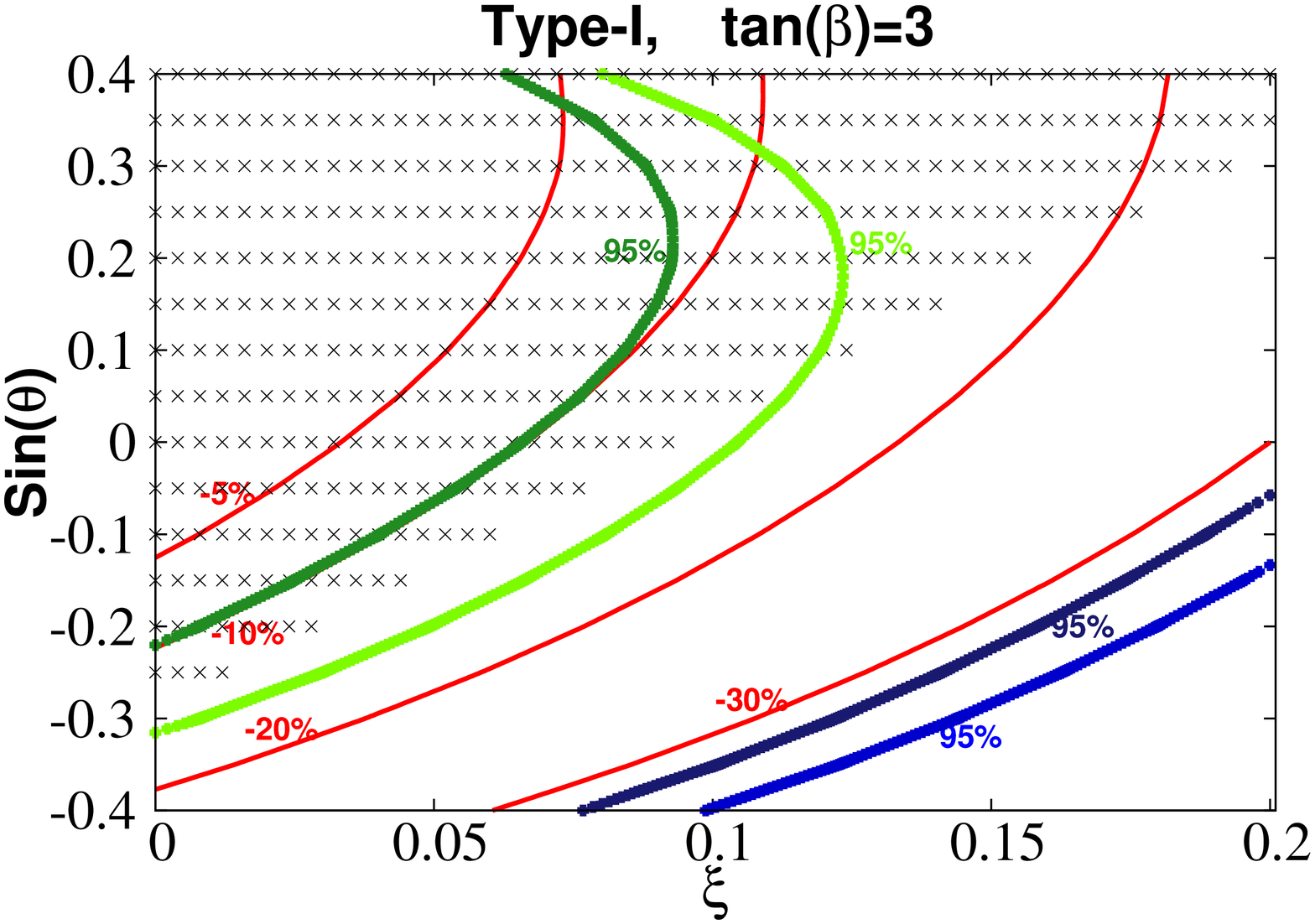}\hspace{2.mm}
\includegraphics[width=0.22\linewidth,angle=270]{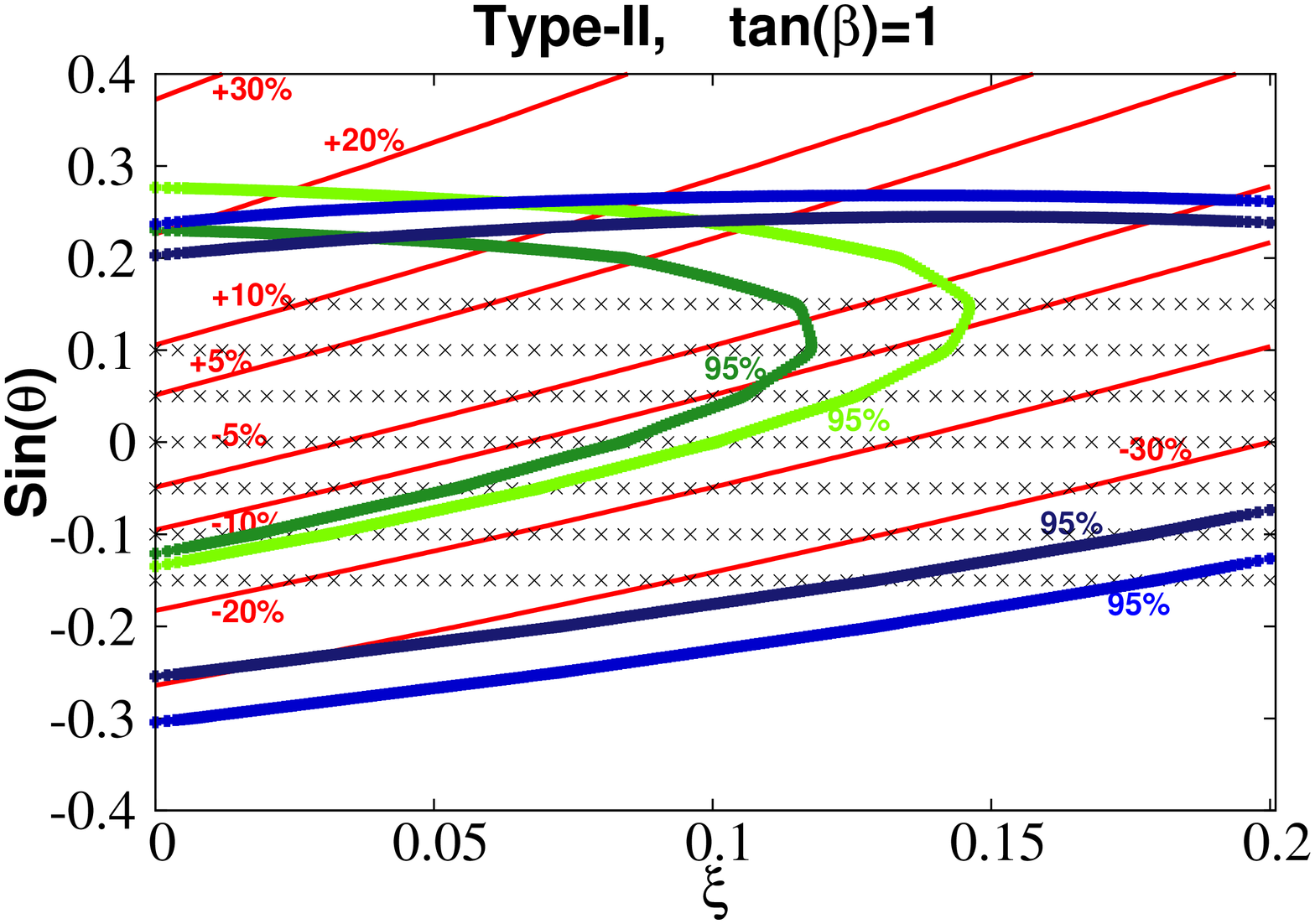}\hspace{2.mm}
\includegraphics[width=0.22\linewidth,angle=270]{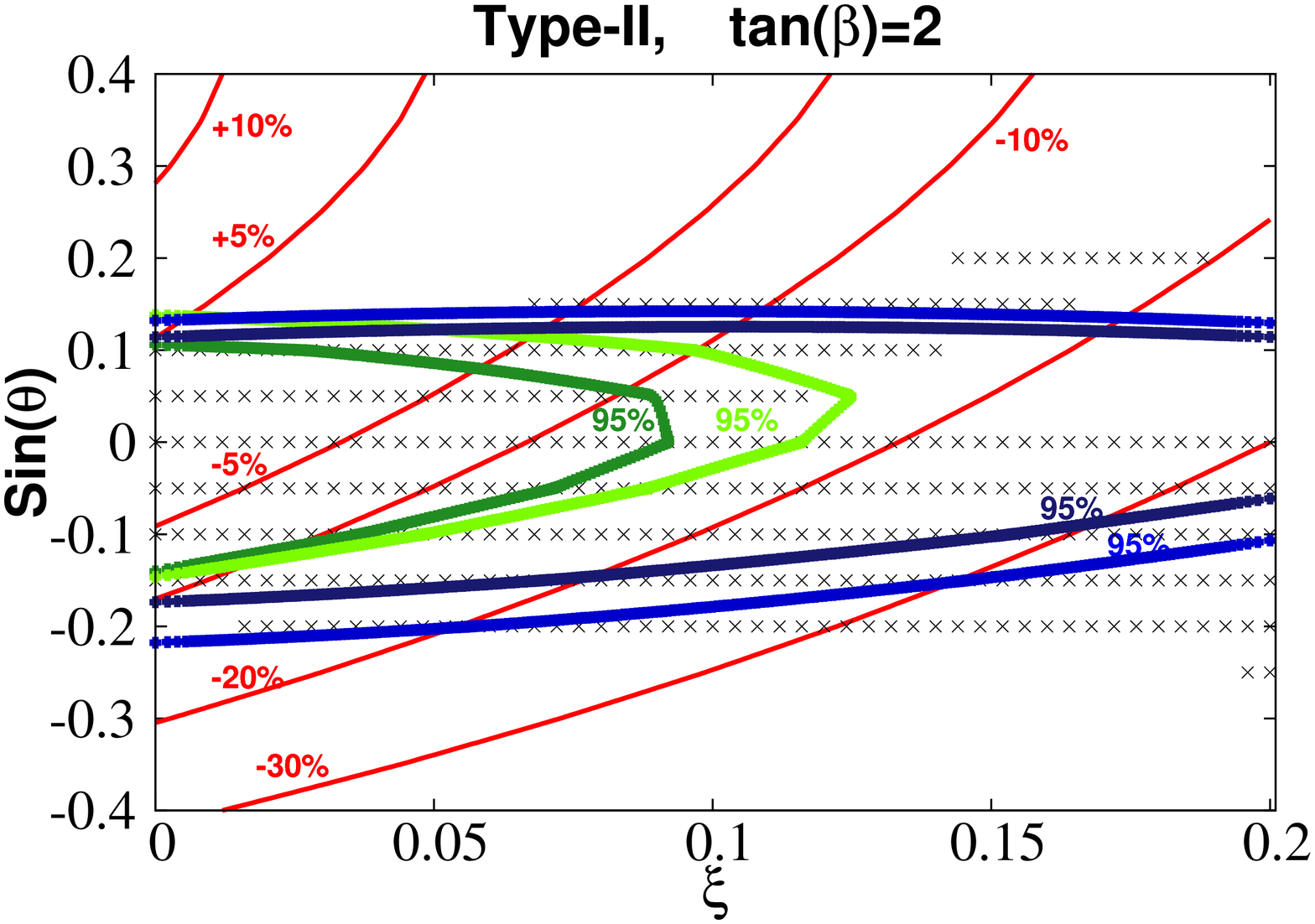}\hspace{2.mm}
\includegraphics[width=0.22\linewidth,angle=270]{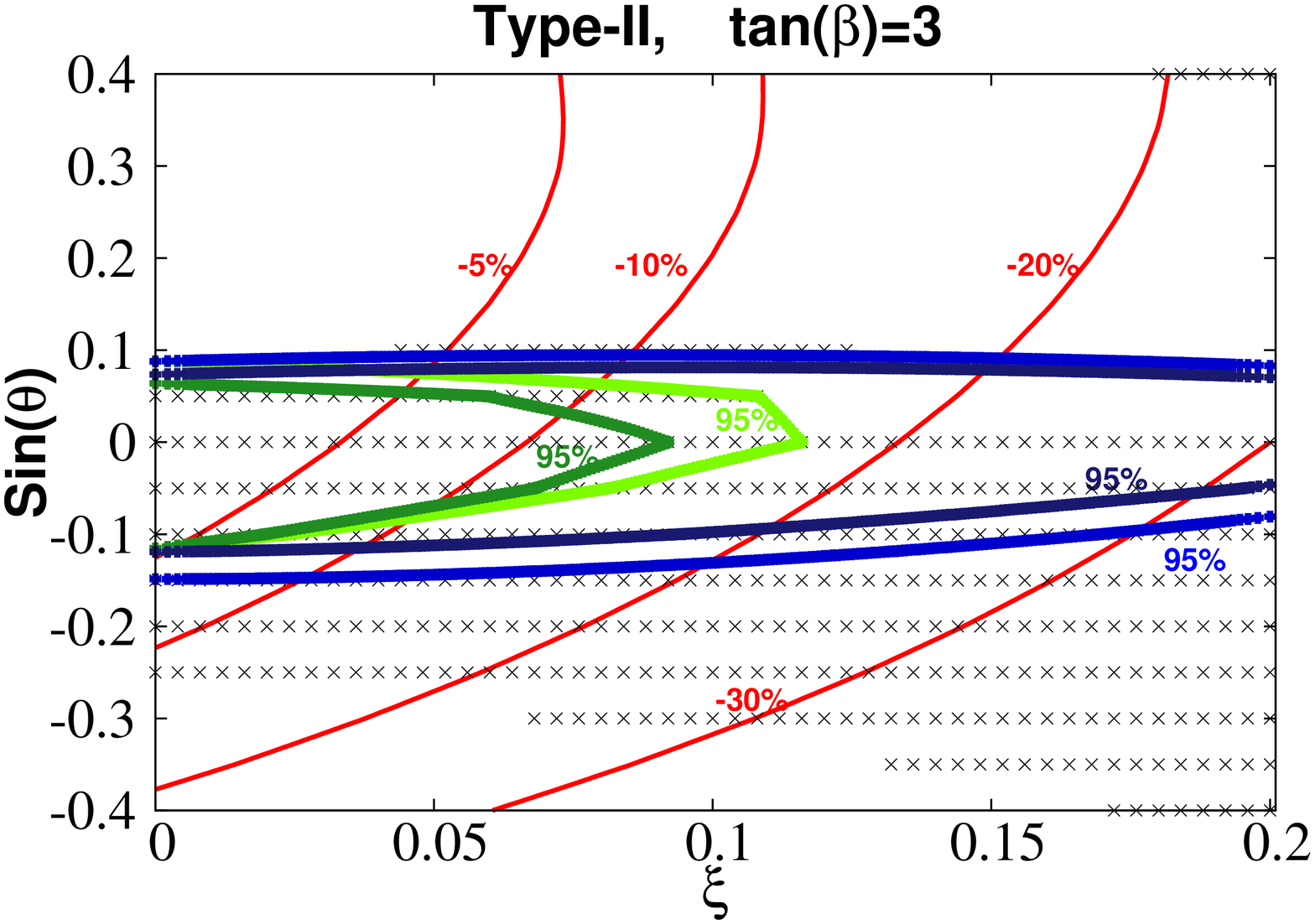}\hspace{2.mm}
\includegraphics[width=0.22\linewidth,angle=270]{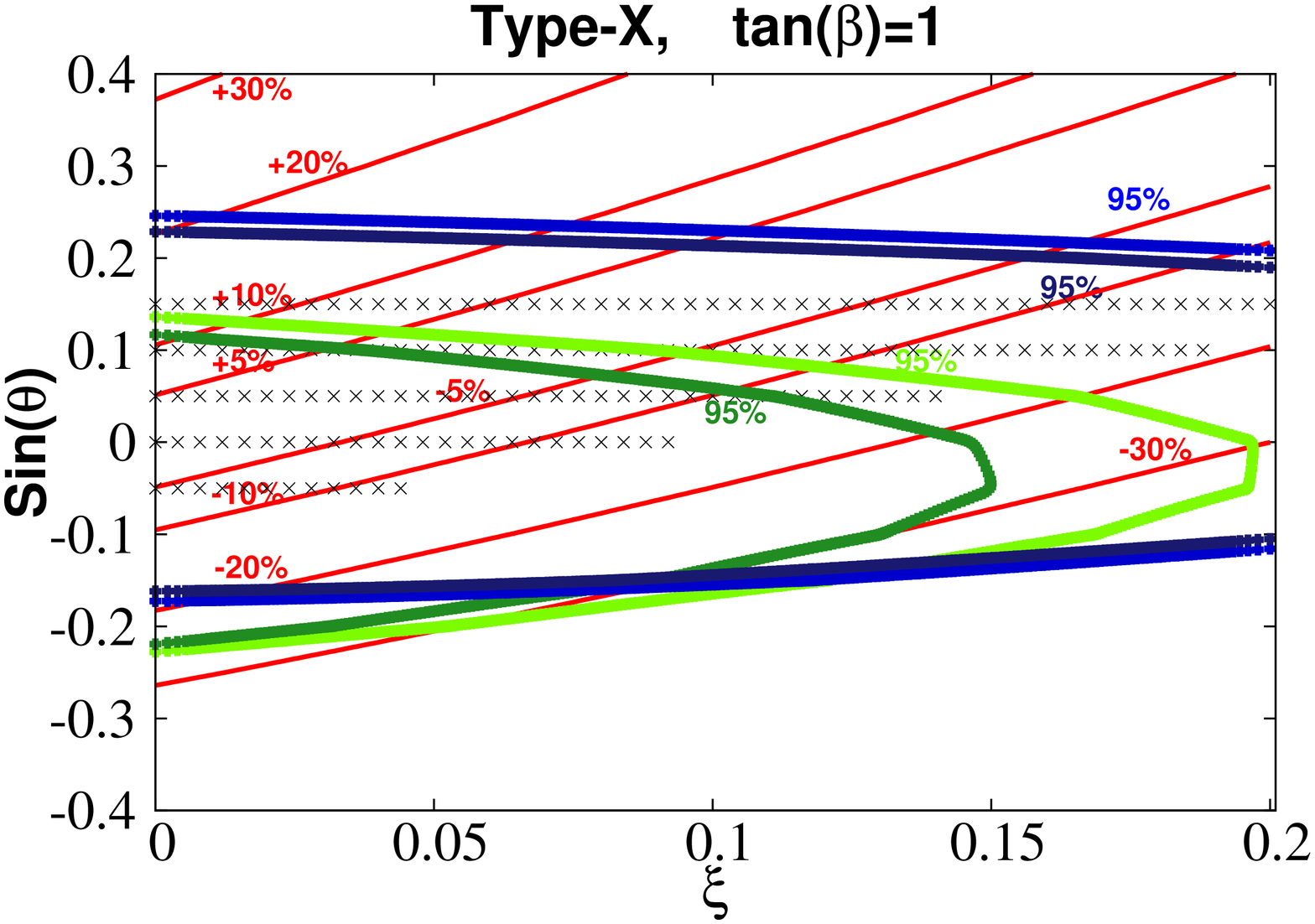}\hspace{2.mm}
\includegraphics[width=0.22\linewidth,angle=270]{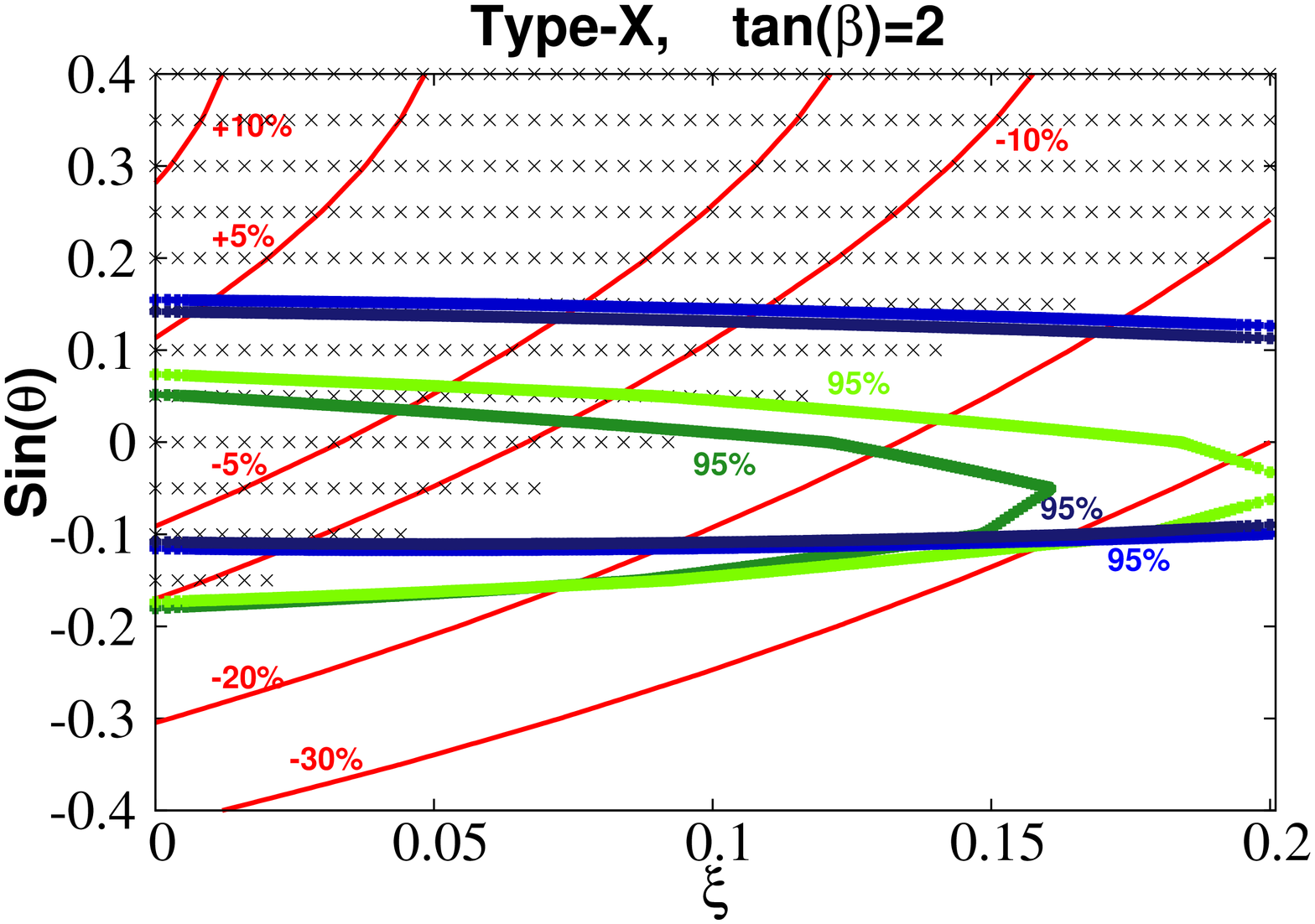}\hspace{2.mm}
\includegraphics[width=0.22\linewidth,angle=270]{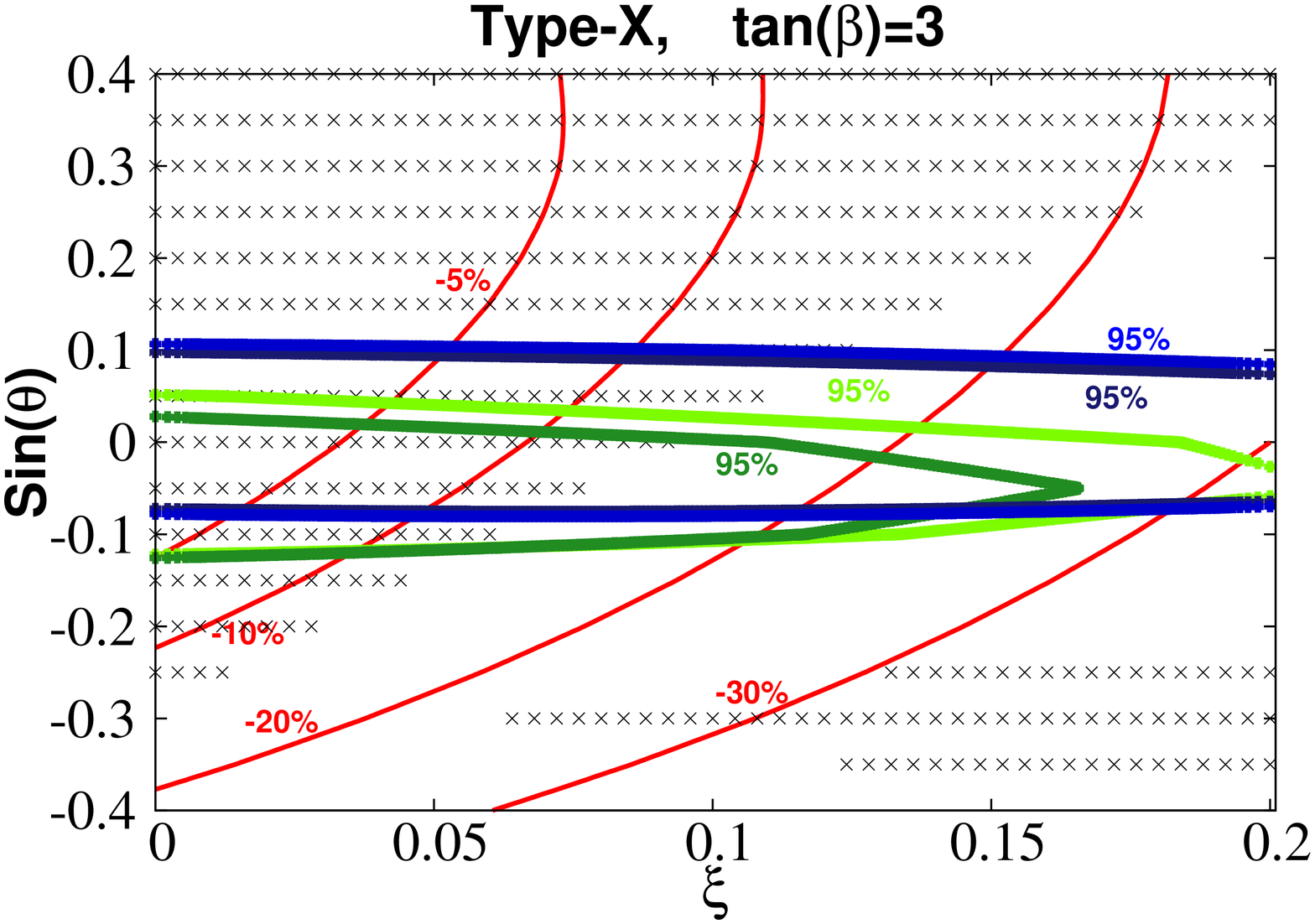}\hspace{2.mm}
\includegraphics[width=0.22\linewidth,angle=270]{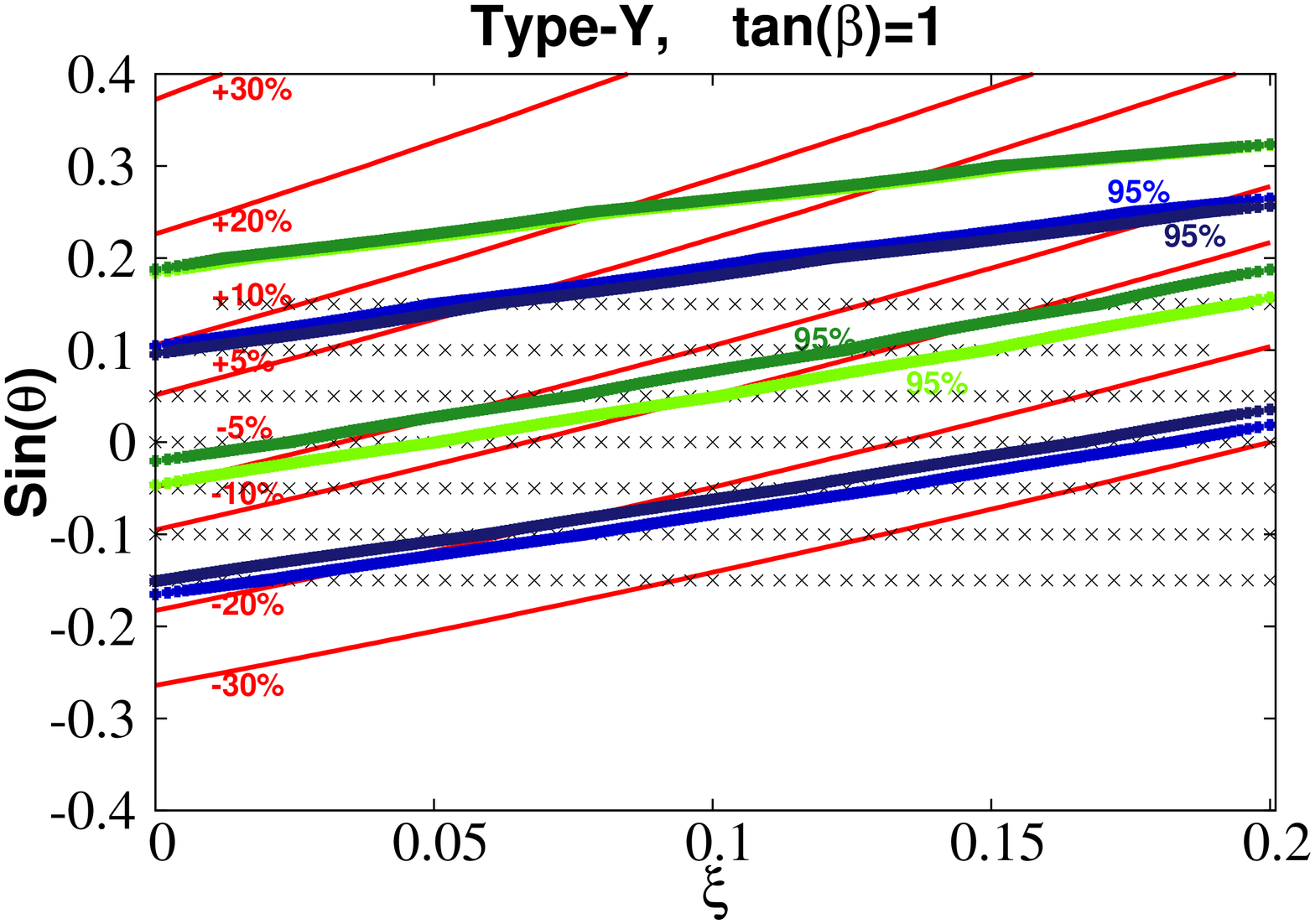}\hspace{2.mm}
\includegraphics[width=0.22\linewidth,angle=270]{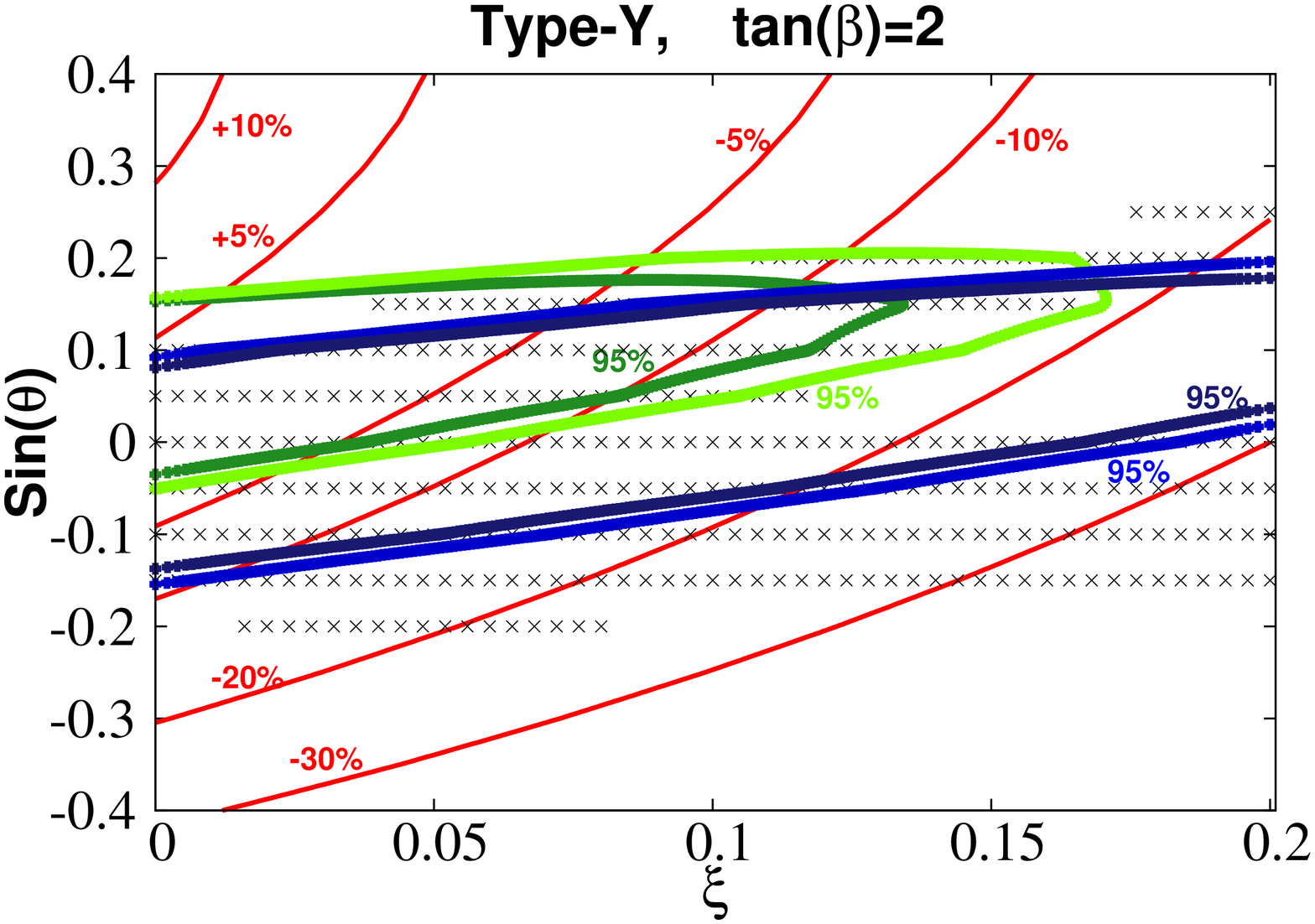}\hspace{2.mm}
\includegraphics[width=0.22\linewidth,angle=270]{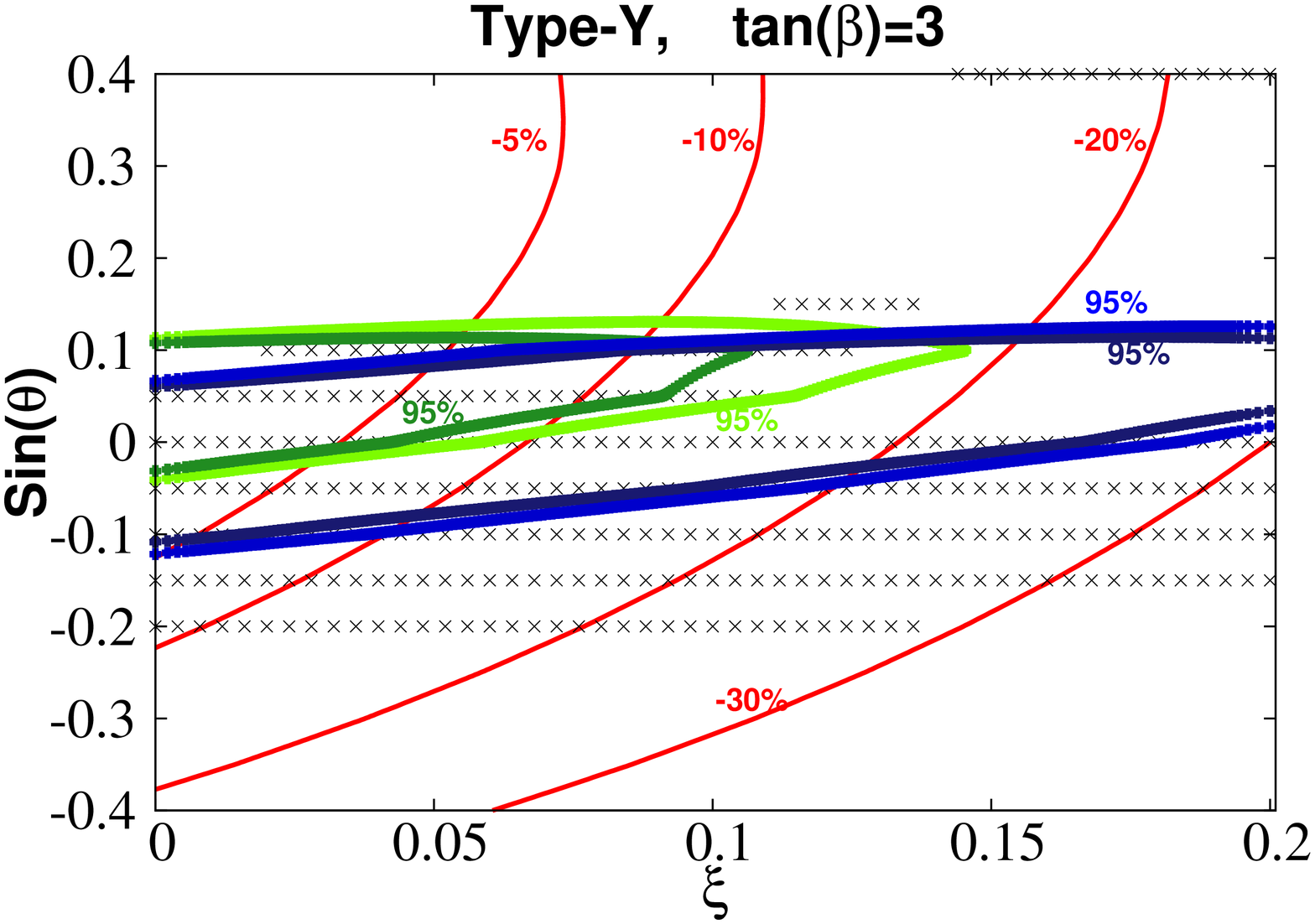}\\ \vspace{4mm}
\caption{Same as Fig.~\ref{HB/HS}, but the red curves are contours of $\Delta\kappa_{t}=y_{htt}/y_{h tt}^{\rm SM}-1$. } 
  \label{HB/HS2}
 \end{center}
 \end{figure}

Before proceeding to study the discussed production modes, we ought to establish the parameter space available to C2HDMs. 
{We first consider constraints coming from  theoretical arguments, namely,  perturbative unitarity and  vacuum stability bounds. 
In Ref.~\cite{DeCurtis:2016scv}, all the eigenvalues of the $s$-wave amplitude matrix for elastic scatterings of two (pseudo)scalar to two (pseudo)scalar processes 
have been derived up to ${\cal O}(s^0)$ terms in the C2HDM case.
Differently from the E2HDM, there is a $s\xi$ dependence in the eigenvalues, so that we need to specify the collision energy $\sqrt{s}$ in addition to the other 
parameters. 
For  vacuum stability bounds, we apply  the same formulae given in the E2HDM~\cite{stability}, i.e., 
\begin{align}
\lambda_1 >0, \quad \lambda_2>0,\quad \sqrt{\lambda_1\lambda_2} + \lambda_3 + \text{min}(0,\lambda_4\pm \lambda_5) > 0. 
\end{align}
In Fig.~\ref{uni}, we show the upper limit on $m_A^{}$ ($=m_H^{}=m_{H^\pm}$) as a function of $\sin\theta$ for a fixed value of $\tan\beta=1$ (left), 2 (center) and 3 (right). 
The black, blue and red curves  show the case for $\xi=0$, 0.04 and 0.08, respectively. 
To calculate the unitarity bound, we take $\sqrt{s}= 1000$ GeV. 
We can see that the allowed region strongly depends upon the choice of the ratio $M/m_A^{}$, which is taken to be 1, 0.8 and 0.6. 
For example, when we take $m_A^{} = 500$ GeV, $M/m_A^{}=0.8$ and $\xi=0$, 
the allowed region of $\sin\theta$ is obtained to be $-0.20\lesssim \sin\theta \lesssim +0.20$ (for $\tan\beta=1$). 
$-0.19\lesssim \sin\theta \lesssim +0.14$ (for $\tan\beta=2$) and $-0.19\lesssim \sin\theta \lesssim -0.18$ (for $\tan\beta=3$). 
If we take a larger value of $\xi$, then 
the unitarity bound is slightly relaxed as compared to the case  $\xi=0$, as it was already mentioned in Ref.~\cite{DeCurtis:2016scv}, while 
the vacuum stability bound  does not change significantly. 
We note that the case of $M \gtrsim m_A^{}$ is highly disfavored by the vacuum stability bound, which induces us
to settle on the illustrative relation $M=0.8  m_A^{}$ ($=m_H^{}=m_{H^\pm}$) for the remainder of the paper. 
}

{Next, we discuss the  constraints from collider experiments. 
We consider the situation which will emerge  at the end of the LHC era 
concerning the investigation of the already discovered 125 GeV Higgs state and that of potential additional signals from an extended Higgs sector.}
Herein, we assume that no additional Higgs states will have been discovered by the LHC, neither in standard luminosity conditions (i.e., after
300 fb$^{-1}$) nor in high luminosity ones (i.e., after 3000 fb$^{-1}$). Even so, we need to decide whether, after such
luminosity values will have been accrued, the measurements of the couplings to SM objects of the Higgs  discovered state will
be as at present (with, of course, a higher precision deriving from the higher statistics) or different. 
In this respect, we will here consider two possible scenarios.  On the one hand, we will assume that current central values of such measurements will have been confirmed. On the other hand, we will assume that SM central values will eventually have been established.  We will show that, under either condition, a future $e^+e^-$ collider will be in a position to disentangle a C2HDM from an E2HDM, through the study of single- and double-$h$ production modes.

In Figs.~\ref{HB/HS}--\ref{HB/HS2} we compare the Higgs sector predictions in C2HDMs (and their E2HDM limits) with the existing exclusion bounds  from the LHC experiments at 95\% CL via the {HiggsBounds} tool  (v4.3.1)  \cite{HB1,HB2,HB3,HB4,HB5,HB6} on the  $(\xi, \sin\theta) $ plane. We consider the usual  Yukawa Type-I, -II, -X and -Y configurations for $\tan\beta=1,2,3$.   Also a  $ \Delta\chi^{2} $  evaluation is made via the {HiggsSignal} package  (v1.4.0)  ~\cite{HS}  to obtain compatibility of the projected exclusion limits from measurements of signal strengths
assuming the (currently) {measured central values} (green), as reported in Tab. \ref{channels} in the Appendix
 from  ATLAS data\footnote{We could not use CMS data in this extrapolation as the public sources that we could access did not report the statistic and systematic errors on the measurements separately.}, as well as the {SM ones} (blue),  with the E2HDMs ($\xi=0$) and C2HDMs ($\xi\ne0$) predictions after  300 fb$ ^{-1} $   (light-green and -blue curves) and 3000 fb$ ^{-1} $  (dark-green and -blue curves) of  LHC luminosity. Herein, the green and  blue contours  present the $ \Delta\chi^{2} $=6.18 (95.45\% CL)  regions. 
To be specific, we summarise in Tab.~\ref{sin-limit} the bounds on $\sin\theta$  in a Type-I, -II, -X and -Y C2HDM obtained by  performing a 95\% CL  fit  using the measurements  listed in Tab.~\ref{channels} in the Appendix  with statistical errors rescaled to an integrated luminosity of 3000 fb$^{-1}$.  Here we consider $ \xi=0,~0.04,~0.08$  (the first one corresponding to the E2HDM limit) and $\tan\beta=1,2,3$. 
\begin{table}[!t]
\resizebox{\columnwidth}{!}{
\scalebox{0.12}{
\begin{tabular}{ll}
\begin{tabular}{|c|c|c|c|c|c|}\hline
$ \xi $ &  $ \tan\beta=1 $ &$ \tan\beta=2 $& $ \tan\beta=3 $\\ \hline
$ \xi=0 $ & \makecell{Type-I ~$-0.05\le s_{\theta} \le 0.15$\\
Type-II ~$ -0.12\le s_{\theta} \le 0.10$\\ Type-X~ $ -0.05\le s_{\theta} \le 0.12$\\ Type-Y~ $ -0.02\le s_{\theta} \le 0.10$}& \makecell{Type-I ~ $ -0.14\le s_{\theta }\le 0.52$\\
Type-II ~$-0.14\le  s_{\theta }\le 0.10$ \\ Type-X $ -0.15\le s_{\theta }\le 0.05$\\ Type-Y~ $ -0.04\le s_{\theta }\le 0.10$}& \makecell{Type-I ~ $-0.22\le s_{\theta} \le 0.51$ \\
Type-II ~$ -0.12\le s_{\theta }\le 0.05$ \\ Type-X~ $ -0.13\le s_{\theta}\le 0.03$\\ Type-Y~ $ -0.03\le s_{\theta }\le 0.05$} \\ \hline

$ \xi=0.04 $ & \makecell{Type-I ~$ 0.00\le s_{\theta }\le 0.15$ \\
Type-II ~$ -0.07\le s_{\theta }\le 0.15$ \\ Type-X ~$ -0.05\le s_{\theta}\le  0.10$\\ Type-Y ~$ 0.00\le s_{\theta }\le 0.15$}& \makecell{Type-I  ~$ -0.05\le s_{\theta }\le 0.47$\\
Type-II ~$ -0.10\le s_{\theta }\le 0.09$ \\ Type-X ~$ -0.10\le s_{\theta }\le 0.04$\\ Type-Y ~$ 0.00\le s_{\theta }\le 0.15$}& \makecell{Type-I  ~$ -0.10\le s_{\theta }\le 0.45$ \\
Type-II ~$ -0.08\le s_{\theta }\le 0.05$\\ Type-X ~$-0.12\le  s_{\theta }\le 0.02$\\ Type-Y ~$ 0.00\le s_{\theta }\le 0.10$} \\ \hline

$ \xi=0.08 $ & \makecell{Type-I ~$ 0.06\le s_{\theta }\le 0.15$\\
Type-II ~$ -0.01\le s_{\theta }\le 0.15$  \\ Type-X ~$0.00 \le s_{\theta }\le 0.07$\\ Type-Y ~$ 0.06\le s_{\theta }\le 0.15$}& \makecell{Type-I  ~$ 0.08\le s_{\theta }\le 0.39$ \\
Type-II ~$ -0.03\le s_{\theta }\le 0.06$ \\ Type-X ~$ 0.00\le  s_{\theta }\le 0.02$\\ Type-Y ~$ 0.05\le s_{\theta }\le 0.15$}& \makecell{Type-I  ~$0.08\le s_{\theta }\le 0.34$ \\
Type-II ~$ -0.03\le s_{\theta }\le 0.02$ \\ Type-X ~$ 0.00\le s_{\theta }\le 0.01$\\ Type-Y ~$ 0.04\le s_{\theta }\le 0.10$ } \\ \hline
\end{tabular}
\end{tabular}
}}
\vspace{5mm}
\caption{Allowed values of $\sin\theta$  in a Type-I, -II,-X and -Y C2HDM with fixed values of $\xi$ and $\tan\beta$ 
obtained by performing both a 95\% CL fit using the measurements listed in Tab.~\ref{channels} in the Appendix and existing data at the LHC. 
For the former fit, statistical errors are rescaled to an integrated luminosity of 3000 fb$^{-1}$. Here we take $m_H^{}=m_A^{}=m_{H^\pm}=500$ GeV and $M=0.8 m_A$. 
}
\label{sin-limit}
\end{table}

Typically, more parameter space with positive values of $ \sin\theta $ enters the  95\% CL contours of the Type-I C2HDM, compared to those of  the Type-II, -X and -Y C2HDMs, for both luminosity data sets.  Overall, 95\% CL contours obtained
under the assumption of  SM central values for the Higgs signal strength measurements (blue curves) enclose larger parameter regions compared to  those obtained
adopting  (currently) measured  central values of the latter  (green curves),  for all Yukawa types.
({The experimental channels excluding most parameter regions in Figs.~\ref{HB/HS}--\ref{HB/HS2} are listed in  Tab.~\ref{exc-channels} in the Appendix.)}

In Figs.~\ref{HB/HS}-\ref{HB/HS2} we also show (in red) the contours of $\kappa_V^{}$ and $\kappa_t$ defined by 
$\kappa_V^{}= g_{hVV}^{}/g_{hVV}^{\text{SM}}$ and $\kappa_t^{}= y_{htt}^{}/y_{htt}^{\text{SM}}$. These red lines 
identify the coupling deviations possible in the relevant C2HDM (with respect to the SM)
as a function of $\sin\theta$ and $\xi$ (plus $\tan\beta$ for the Yukawa coupling) 
that we intend to probe using the described single- and double-$h$ production modes, within the limits imposed by Tab.~\ref{sin-limit}. Note 
that the latter collects the most stringent bounds possible at 95\% CL following the High Luminosity LHC (HL-LHC) runs (i.e., after 3000 fb$^{-1}$ of luminosity).   

\section{Single Higgs Boson Production}

\begin{figure}[t]
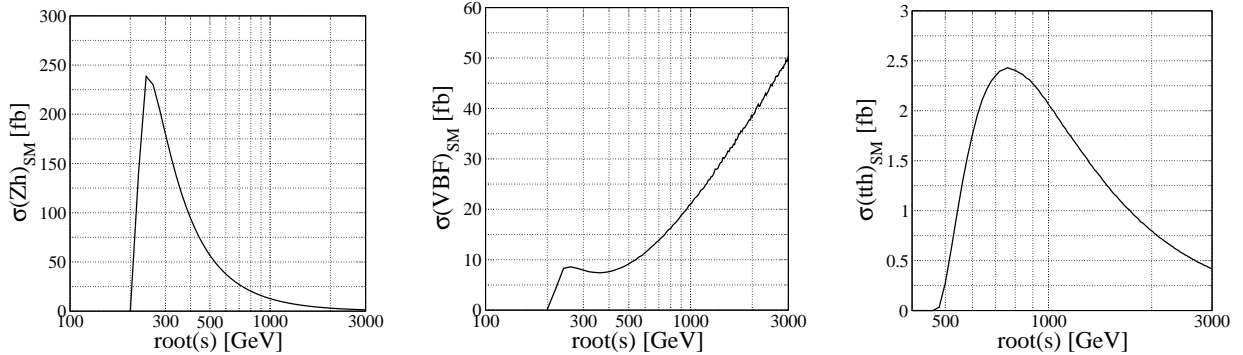

\begin{center}
\includegraphics[width=50mm]{Zh_sm.eps}\hspace{5mm}
\includegraphics[width=50mm]{VBF_sm.eps}\hspace{5mm}
\includegraphics[width=50mm]{tth_sm.eps}
\caption{Cross sections for $e^+e^-\to Zh$ (left),  $e^+e^-\to e^+e^-h$ (center) and 
$e^+e^-\to t\bar{t}h$ (right) processes as  functions of $\sqrt{s}$ in the SM. }
\label{cross_h}
\end{center}
\end{figure}

In this section, we discuss the three single-Higgs boson production processes, namely, Higgs-Strahlung (HS), Vector Boson Fusion (VBF) and associated production with top quarks 
($t\bar{t}h$). 
We calculate all the cross sections for  the Type-I Yukawa interaction, but, as already stressed, for our $\tan\beta$ choices, the results are also valid for the other Yukawa types. 

The reference SM cross sections for these processes
are found in Fig.~\ref{cross_h}, as a function of the Centre-of-Mass (CM) energy of the collider $\sqrt{s}$.  
Here, we can see that the HS and  $t\bar{t}h$ production cross sections can be maximal at just above their thresholds, i.e., 
$\sqrt{s} \sim 215$ and $\sim 425$ GeV, respectively. 
When $\sqrt{s}$ gets larger, the cross sections monotonically decrease because of the $s$-channel topology in both cases. 
In contrast, for the VBF process, the cross section increases according to $\log \sqrt{s}$ due to the $t$-channel topology. 
Recall that precision on such cross sections at future $e^+e^-$ colliders, quite irrespectively of the machine configuration and energy, is expected to be at the percent level or even less
(especially for HS and VBF).

Let us first consider  the HS and VBF processes, where,  in both E2HDMs and C2HDMs, there is only one diagram containing the $hZZ$ vertex, just like in the SM. 
Thus, these production cross sections are simply obtained from the corresponding ones in the SM upon multiplying for 
the squared scaling factor of the $hZZ$ coupling $\kappa_V^2$. 
Therefore, by measuring these cross sections, one can extract $\kappa_V^2$. 

\begin{figure}[t]
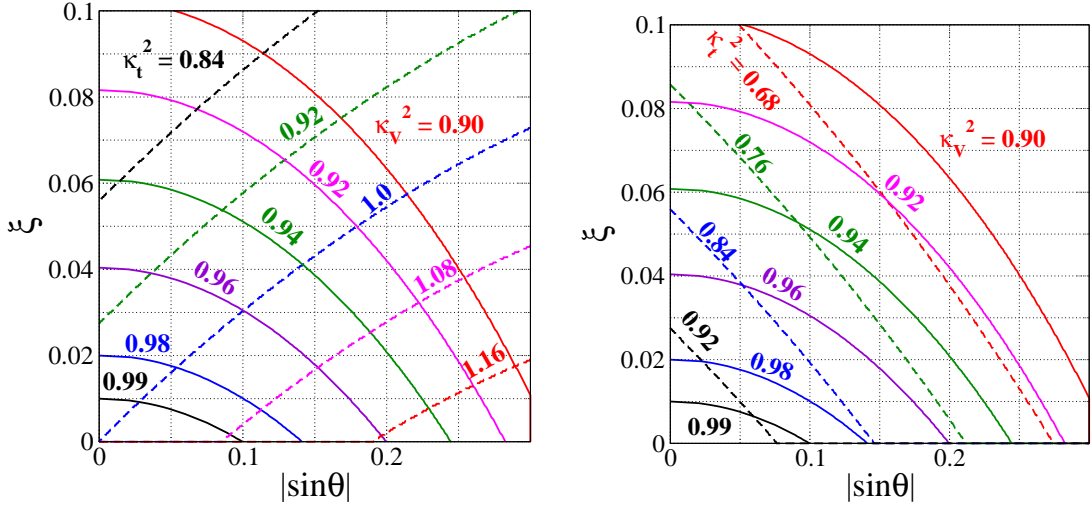

\begin{center}
\includegraphics[width=70mm]{kvsq-contour_p.eps}\hspace{5mm}
\includegraphics[width=68mm]{kvsq-contour_m.eps}
\caption{Contour plots of $\kappa_V^2= (g_{hVV}/g_{hVV}^{\text{SM}})^2$ (solid)
and $\kappa_t^2= (y_{htt}/y_{htt}^{\text{SM}})^2$ (dashed) in the $(|\sin\theta|,\xi)$ plane for 
$\sin\theta<0$ (left) and
$\sin\theta>0$ (right). Contour plots of $\kappa_t^2$ are for $\tan\beta=2$.
 }
\label{fig_hVV}
\end{center}
\end{figure}

In all C2HDM types, $\kappa_V^{}$ depends on two parameters, $|\sin\theta|$ and $\xi$, as we can see  in Eq.~(\ref{gauge}), while in the corresponding E2HDM cases $(\xi=0)$ only
one parameter is involved. 
This means that, if $\kappa_V^{}$ is measured at $e^+e^-$ colliders, this determines the allowed combinations of $
|\sin\theta|$ and $\xi$ via Eq.~(\ref{gauge}). 
In Fig.~\ref{fig_hVV}, we show the contour of $\kappa_V^2$ on the $(|\sin\theta|,\xi)$ plane. 
For example, if  $\kappa_V^2=0.94$ (the green solid curve), 
the value of $|\sin\theta|$ is determined to be  about $0.245$ at $\xi=0$, which corresponds to the E2HDM case, 
while this can vary from 0 to 0.245 in the C2HDM one by varying $\xi$  from about 0.06 to 0. 
Therefore, in order to disentangle $\xi$ and $|\sin\theta|$ in C2HDMs, we need  further inputs from experiment. In particular,  
once $\kappa_V$ is fixed, one can then predict the deviations expected in $\kappa_t$ by fixing the sign of $\sin\theta$ and the value of  
$\tan\beta$. This way, in fact, one can get $\kappa_t$ from Eq.~(\ref{Yukawas}). The $\kappa_t$ contours are also shown in  Fig.~\ref{fig_hVV}. 
In contrast, in the E2HDM case, once $\kappa_V$ and $\tan\beta$ are known, only two values of $\kappa_t$, depending on the sign of $\sin\theta$, are uniquely obtained. This delineates therefore a strategy to follow in
order to possibly separate the two assumptions of 2HDMs, elementary and composite, i.e., via the simultaneous extraction of $\kappa_V$ from HS and VBF 
and measurement of the event rates for associated production with top quarks. In short, at fixed $\kappa_V^2$, there could well be values of the $t\bar th$ cross section obtainable in C2HDMs which are instead unattainable in E2HDMs.

\begin{figure}[t]
\begin{center}
\includegraphics[width=150mm]{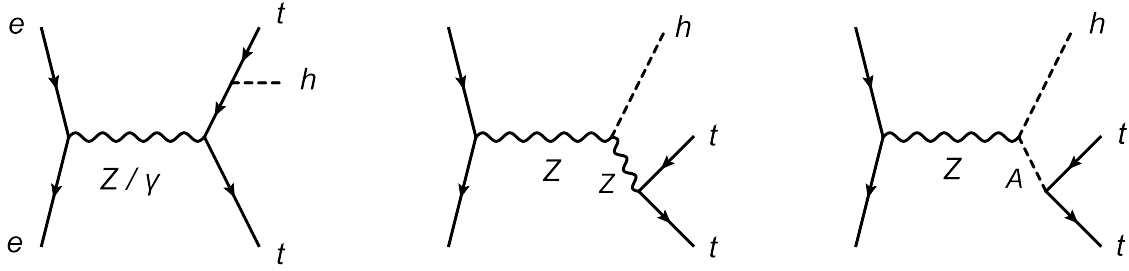}
\caption{Representative Feynman diagrams for the $e^+e^- \to t\bar t h$ process. }
\label{diag_tth}
\end{center}
\end{figure}

Before doing so though,   let us investigate the $t\bar{t}h$ production cross section as a function of $\sin\theta$,  $\xi$ and $\tan\beta$. 
Contrary to the HS and VBF cases, this process requires a more involved treatment. 
As clear from Fig.~\ref{diag_tth}, there are three representative diagrams entering such a process, namely: 
(i) $e^+e^-\to t \bar t$ production followed by $h$ emission from $t$ and $\bar t$ (first diagram of Fig.~\ref{diag_tth}), 
(ii) $e^+e^-\to Z^{*} h$ production followed by $Z^{*}\to t\bar t$ (second diagram of Fig.~\ref{diag_tth}) and 
(iii) $e^+e^-\to A^{(*)} h$ production followed by $A^{(*)}\to t\bar t$  (third diagram of Fig.~\ref{diag_tth}). 
By looking at the functional form of $g_{hVV}$ and $ g_{AhZ} $ in Eq. (\ref{gauge})
and $y_{htt}$ and $ {\tilde y}_{Att} $ in Eq. (\ref{Yukawas}), it is clear that the cross section for $e^+e^-\to t\bar t h$ 
does not scale trivially with respect to the SM in neither the E2HDM nor the C2HDM case. 

\begin{figure}[t]
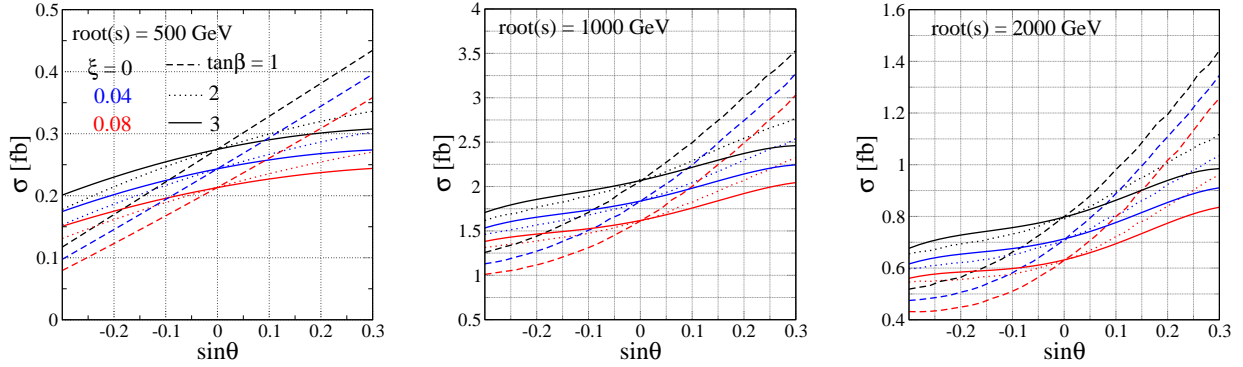

\begin{center}
\includegraphics[width=50mm]{tth_rs500_m320.eps}\hspace{5mm}
\includegraphics[width=50mm]{tth_rs1000_m400.eps}\hspace{5mm}
\includegraphics[width=50mm]{tth_rs2000_m400.eps}
\caption{Cross section for the $e^+e^-\to t\bar{t}h$ process as a function of $\sin\theta$. 
The value of $\tan\beta$ is taken to be 1 (dashed), 2 (dotted) and 3 (solid), and that of $\xi$ is taken to be $0$ (black), 0.04 (blue) and 0.08 (red). 
The collision energy and the mass of $A$ $(\sqrt{s},m_A^{})$ is taken to be (500, 400), (1000, 500) and (2000,500) in GeV unit for the left, center and right panels, respectively.  
}
\label{cross_tth}
\end{center}
\end{figure}

In Fig.~\ref{cross_tth}, we show the cross section for the $t\bar{t}h$ production process as a function of $\sin\theta$ for several fixed values of $\tan\beta$ and $\xi$. 
The collision energy $\sqrt{s}$ is taken to be 500 (left), 1000 (center) and 2000 GeV (right). 
We can see that the cross section gets smaller when we take a smaller value of $\sin\theta$. 
In addition, when we take $\sqrt{s}$ to be 1 TeV or 2 TeV, the on-shell production of $A$ is realised. This significantly enhances the cross section 
as it is seen by comparing the case for $\sqrt{s}=500$ and 1000 or 2000 GeV. 
Concerning the differences between the E2HDM and C2HDM cases, parametrised by $\xi$, we find that the cross section is smaller for larger values of $\xi$.
It is also seen that the ratio of the cross section with $\xi = 0$ to that with $\xi=0.04$ (or 0.08) for a fixed value of $\tan\beta$, $\sin\theta$ and $\sqrt{s}$
does not depend much on the choice of $\tan\beta$, $\sin\theta$ and $\sqrt{s}$. The $\xi$ dependence acts quite like an overall rescaling of the $t \bar t h$ cross section.
Significant deviations are possible between the E2HDM and C2HDM cases. 
If we compare the two scenarios for the same value of $\sin\theta$,  
the difference is ${\cal O}(20\%)$ or so in all Yukawa types, for $\sin\theta$ and $\xi$ combinations allowed
by Tab.~\ref{sin-limit}. 

However, as intimated, what we really need to compute is the $e^+e^-\to t\bar t h$
cross section for a fixed value of $\kappa_V^2$, as this will promptly be measured at future electron-positron machines via the HS and VBF processes. 
As stressed already, the $t\bar t h$ results are quite independent of the choice of the type of Yukawa interactions. In contrast, as shown in Figs.~\ref{HB/HS}--\ref{HB/HS2}, the bounds from collider data are different for the various 2HDM types. In Tab.~\ref{sin-limit} we have reported the 95\% CL  bounds on $\sin\theta$ after 3000 fb$^{-1}$ of LHC accumulated data, for
fixed $\xi$ and $\tan\beta$. Under the hypothesis of having measured $\kappa^2_V$, this information determines the allowed $\sin\theta$ values that can be related to $\xi$,
as shown in Fig.~\ref{fig_hVV}.

\begin{figure}[t]
\begin{center}
\includegraphics[width=70mm]{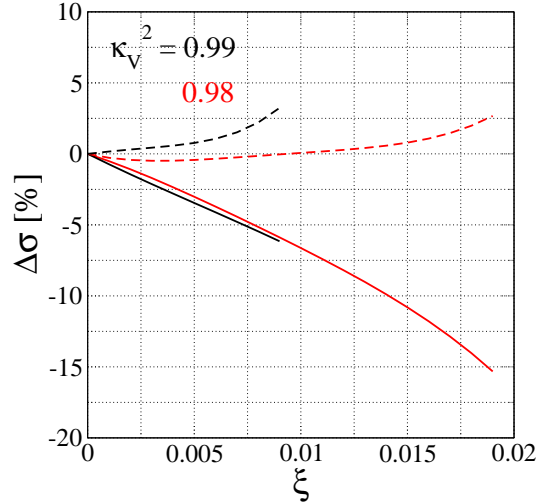}\hspace{5mm}
\caption{Deviations in the cross section $\Delta \sigma \equiv (\sigma_{\text{C2HDM}}/\sigma_{\text{E2HDM}} -1)$ for the $e^+e^-\to t\bar{t}h$ process 
at fixed $\kappa_V^2=0.99$ (black curves) and 0.98 (red curves).  We take $\tan\beta =2$ and $\sqrt{s}=1000$ GeV. 
For each $\kappa_V^2$, we show the case of $\sin\theta>0$ (solid lines) and $\sin\theta<0$ (dashed lines). 
We only show the result allowed by the unitarity and vacuum stability bounds and by the future LHC data assuming 3000 fb$^{-1}$ with 95\% CL. 
The latter bound is for the Type-I C2HDM. }
\label{dcross_tth}
\end{center}
\end{figure}

In Fig.~\ref{dcross_tth}, we therefore show the deviation in the $t \bar t h$ cross section in C2HDMs relative to E2HDMs, i.e., 
$\Delta \sigma\equiv (\sigma_{\text{C2HDM}}/\sigma_{\text{E2HDM}} - 1)$, 
as a function of $\xi$ for  two values of $\kappa_V^2$ = 0.99 and 0.98. 
We here take   $\sqrt{s}= 1000$ GeV,  $m_A^{}=500$ GeV and $\tan\beta = 2$.
Now, $|\sin\theta|$ is determined for each fixed value of $\xi$ (see Fig.~\ref{fig_hVV}). 
Since the sign of $\sin\theta$ cannot be determined by measuring $\kappa_V^2$, we show the cases for $\sin\theta >0$ (solid curves) and  $\sin\theta < 0$ (dashed curves) separately.
The result is that we can still have a very large difference between the elementary and composite $e^+e^- \to t \bar t h$ cross section. 
For example, if the measured value of $\kappa_V^2$ were 0.98, then $\Delta\sigma$ can reach $-15\%$ for $\sin\theta > 0$.
This behaviour can be explained as follows.  Once $\kappa_V^2$ is specified, this fixes $|\sin\theta|$ in the E2HDM. 
In contrast, in the C2HDM the value of $|\sin\theta|$ can be lower depending on  $\xi$. 
For the case of $\sin\theta > 0$, when the value of $\xi$ gets large, $\sin\theta$ decreases (approaching 0). 
From Fig.~\ref{cross_tth}, it is seen that the cross section becomes small when $\sin\theta$($\xi$) decreases(increases), so that the ratio becomes smaller. 
Conversely, for $\sin\theta< 0$, 
a larger value of $\xi$ corresponds to a larger value of $\sin\theta$, 
so that the reduction of the cross section by a larger value of $\xi$ can be cancelled through a larger value of $\sin\theta$. Either way, large values for $\Delta \sigma$, well beyond the expected precision of the $e^+e^-\to t\bar th$ cross sections can be attained for allowed values of
$\sin\theta$ and $\xi$ (for a given $\tan\beta$). 

In conclusion, we find that, after having enforced theoretical bounds and experimental limits from the high luminosity option of the LHC,  there are parameter space regions of C2HDMs 
predicting  cross sections for $t\bar{t} h$ production that cannot be realised in E2HDMs for a given value of  $\tan\beta$  when 
$\kappa_V^2$ is precisely determined via the HS and VBF processes. 
This also suggests that we can extract the value of $\xi$ from the measurement of the $t\bar{t} h$ yield if  $\tan\beta$ is known through the
study of other observables. In fact, such a parameter can be accessed at the LHC, e.g.,  via the precise measurement of the Yukawa couplings of the $h$ state.
In reality, one may also need to know the values of $m_A$, $\Gamma_A$ and the $At\bar t$ coupling, whenever the third topology in Fig.~\ref{diag_tth}
contributes significantly to the   $e^+e^-\to t\bar{t} h$ 
cross section, e.g., when it is resonant, as is the case in Fig.~\ref{cross_tth}, since herein one has $m_A>2 m_t$. (Recall that the $AhZ$ coupling is fixed by the gauge structure, which is
common to both the elementary and composite Higgs scenarios we are considering, so that, unlike 
$m_A$, $\Gamma_A$ and ${\tilde y}_{Att}$, it is not an independent parameter.) As we are working under the condition (already spelt out in the introduction) that the LHC will have not produced any evidence of additional Higgs
bosons other than the SM-like $h$ state (assumption which is indeed encoded in Figs.~\ref{HB/HS}--\ref{HB/HS2}), access to these additional parameters can be gained through the
study of the $e^+e^-\to hA$ cross section and decay rates, which are promptly accessible at a future $e^+e^-$ collider whenever $\sqrt s> m_h+m_A$.    
Ultimately, knowledge of $m_A$, $\Gamma_A$ and the $At\bar t$ coupling would give access to $\kappa_t$ in C2HDMs, for which large deviations from the E2HDMs counterpart are possible over allowed parameter regions,
at the level  of tens of percent, see  red lines in Fig.~\ref{HB/HS2}.

\section{Double Higgs Boson Production}

\begin{figure}[t]
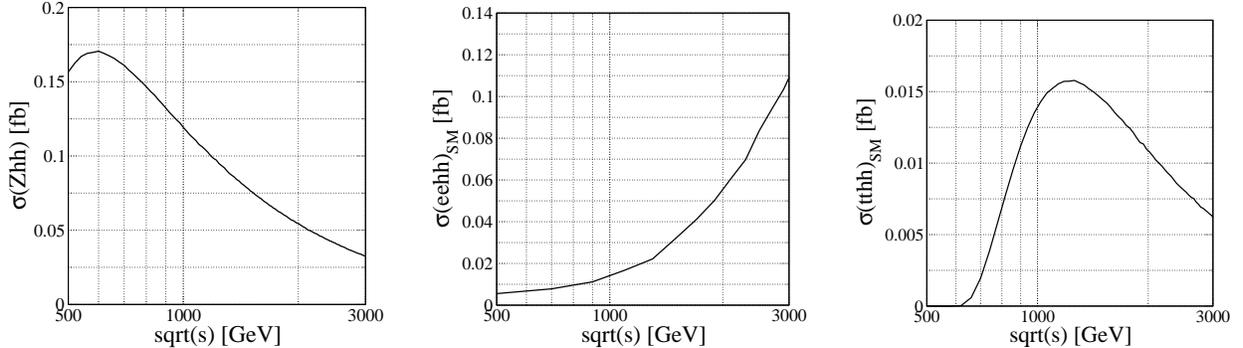

\begin{center}
\includegraphics[width=50mm]{Zhh_sm.eps}\hspace{5mm}
\includegraphics[width=50mm]{eehh_sm.eps}\hspace{5mm}
\includegraphics[width=50mm]{tthh_sm.eps}
\caption{Cross section of the $e^+e^-\to Zhh$ (left),  $e^+e^-\to e^+e^-hh$ (center) and 
$e^+e^-\to t\bar{t}hh$ (right) as a function of $\sqrt{s}$ in the SM. 
}
\label{cross_hh}
\end{center}
\end{figure}

In this section, we tackle the case of double-Higgs production,  
wherein the pair of final state Higgs bosons is made up by two $h$ (SM-like) states. 
The production modes are those already discussed, i.e., HS ($e^+e^-\to Zhh$), VBF ($e^+e^-\to e^+e^- hh$) and associated production with 
top quarks ($e^+e^-\to t\bar t hh$). 
For reference, the cross sections for these processes in the SM are given in Fig.~\ref{cross_hh}.
Typically, in each case, the production cross section is more than hundred times smaller than the corresponding cross section for   single-$h$ production due to the phase space suppression. 
Double-Higgs production enables one to access triple-Higgs self-couplings, specifically, in the case of $h$ pairs, the 
$hhh$ and $Hhh$ vertices. As mentioned earlier, while the constraints that can be extracted on these couplings at the LHC are rather poor, 
with precisions of  ${\cal O}(100\%)$, the accuracy achievable at future $e^+e^-$ colliders can be less
close to 10\%.   

\begin{figure}[t]
\begin{center}
\includegraphics[width=150mm]{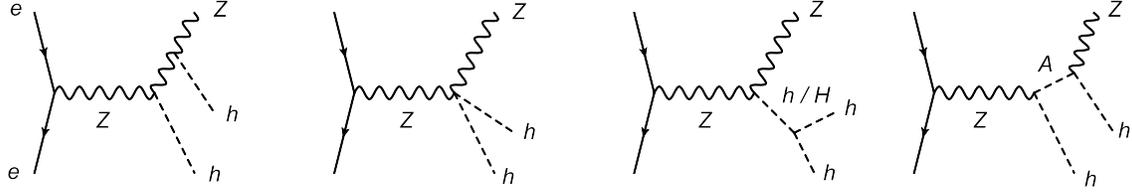}
\caption{Representative Feynman diagram for the $e^+e^- \to Zhh$ process. }
\label{diag_Zhh}
\end{center}
\end{figure}

\begin{figure}[t]
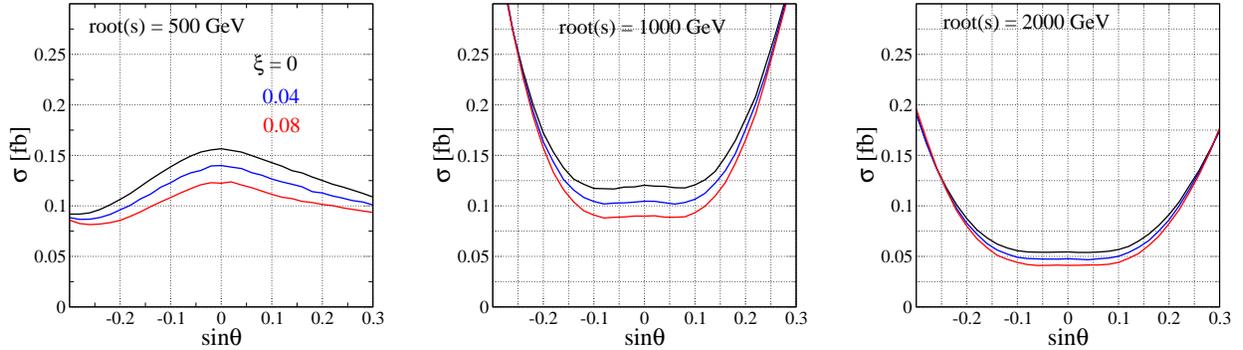

\begin{center}
\includegraphics[width=50mm]{zhh_rs500_m320_tb2.eps}\hspace{5mm}
\includegraphics[width=50mm]{zhh_rs1000_m400_tb2.eps}\hspace{5mm}
\includegraphics[width=50mm]{zhh_rs2000_m400_tb2.eps}
\caption{
Cross section for the $e^+e^-\to Zhh$ process as a function of $\sin\theta$ in the C2HDM with $\tan\beta =2$ and $M = 0.8 m_A^{}$.
The collision energy and the mass $(\sqrt{s},m_A)$ with $m_H^{}=m_A^{}$ is taken to be (500, 400), (1000, 500) and (2000, 500) in GeV unit for the left, center and right panels, respectively.  
}
\label{cross_Zhh1}
\end{center}
\end{figure}

First of all, let us discuss the $Zhh$ production mode. 
The representative Feynman diagrams are shown in Fig.~\ref{diag_Zhh}. 
Differently from the single-$h$ production case, there are here interactions depending on $\lambda_{hhh}$ and $\lambda_{Hhh}$ (in the third diagram), where 
the expressions for these couplings are given in Eqs.~(\ref{hhh}) and (\ref{Hhh}). 
In addition, the fourth diagram contains the propagation of $A$. 
It is important to mention here that the $\tan\beta$ dependence of the cross section for this process only enters via the $\lambda_{hhh}$ and $\lambda_{Hhh}$ couplings and 
their sensitivity to this parameter is very weak for small $\theta$ values. 
In fact, for $\theta\to0$, one has
\begin{align}
\lambda_{hhh} &= -\frac{m_h^2}{2v_{\text{SM}^{}}}\left(1-\frac{\xi}{6}\right) + {\cal O} (\theta^2), \\
\lambda_{Hhh} &= -\frac{\theta}{2v_{\text{SM}^{}}}\left[m_H^2 -2m_h^2 + 4M^2 -\frac{\xi}{6}(m_H^2-2m_h^2-4M^2)
\right] + {\cal O} (\theta^2),
\end{align}
wherein the  $\tan\beta$ dependence only appears at the ${\cal O}(\theta^2)$ level.  

In Fig.~\ref{cross_Zhh1}, we show the production cross section of the $e^+e^- \to Zhh$ process as a function of $\sin\theta$ for three fixed values of $\xi$, i.e, 0, 0.04 and 0.08. 
We here take $\tan\beta = 2$ (indeed, we have checked that the $\tan\beta$ dependence is essentially negligible for $|\sin\theta| \lesssim 0.2$). 
It is seen that the $\sin\theta$ dependence of the cross section is quite different in the case of  $\sqrt{s}=500$ GeV from those when $\sqrt{s}=1000$ and 2000 GeV. 
This can be explained depending upon whether  on-shell $A$ production is possible or not. 
Namely, in the case of $\sqrt{s}=500$ GeV, the diagram including $A$ is not important, because it is off-shell, since $m_A^{}=400$ GeV is larger that $\sqrt s-m_h$. 
In contrast, for the cases with $\sqrt{s}=1000$ and 2000 GeV, a non-zero value of $\sin\theta$ allows one to have on-shell production of both $H$ and $A$ followed by the decays 
$H\to hh$ and $A\to hZ$, respectively, since $\lambda_{Hhh}$ and $g_{AhZ}^{}$ are proportional to $\sin\theta$, as  seen in Eqs.~(\ref{Hhh}) and (\ref{gauge}). 
Therefore, at these two energies, the cross section can be enhanced due to their resonant productions. 
Concerning the $\xi$ dependence, we see that deviations between the C2HDM and E2HDM case remain comparable at all energies, generally being in the 20--30\% range, a result of the interplay between the fact that the aforementioned  $H\to hh$ and $A\to hZ$ decays are not the dominant ones at $m_A=m_H=500$ GeV with the $\xi$ dependence of the $HZZ$ and $AhZ$ couplings.

\begin{figure}[t]
\begin{center}
\includegraphics[width=70mm]{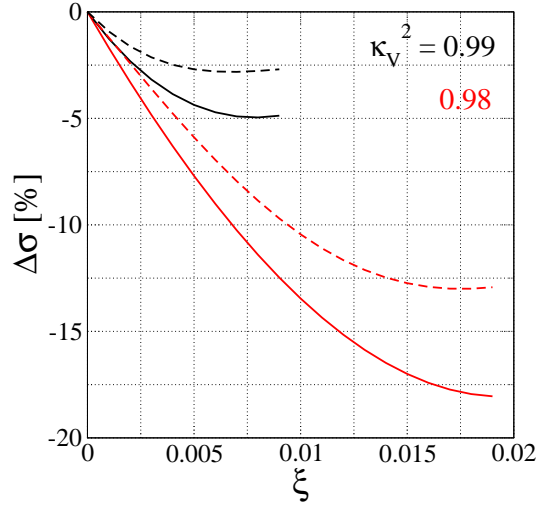}
\caption{Deviations in the cross section $\Delta \sigma \equiv (\sigma_{\text{C2HDM}}/\sigma_{\text{E2HDM}} -1)$ for the $e^+e^-\to Zhh$ process 
at fixed $\kappa_V^2=0.99$ (black curves) and 0.98 (red curves).  
We take $\tan\beta =2$ and $\sqrt{s}=1000$ GeV. 
For each $\kappa_V^2$, we show the case of $\sin\theta>0$ (solid lines) and $\sin\theta<0$ (dashed lines). 
We only show the result allowed by the unitarity and vacuum stability bounds and by the future LHC data assuming 3000 fb$^{-1}$ with 95\% CL
The latter bound is for the Type-I C2HDM. }
\label{dcross_Zhh}
\end{center}
\end{figure}

In Fig.~\ref{dcross_Zhh}, we show the deviations in the $e^+e^-\to Zhh$ cross section from the E2HDM  case appearing
in the C2HDM one 
by considering fixed values of $\kappa_V^2$ = 0.99 and 0.98. 
We here take $\sqrt{s}=1000$ GeV, $m_H^{}=m_A^{}=500$ GeV, $M = 0.8 m_A^{}$ and $\tan\beta=2$.  
We see that negative deviations up to about $-18\%$ seen in the case of $\kappa_V^2=0.98$ and $\sin\theta > 0$
are predicted also after enforcing the bounds from the high luminosity data from the LHC.
The main reason for this is a decreasing $|\sin\theta|$ as $\xi$  gets larger. 
As we explained  above, this result is nearly independent of the choice of $\tan\beta$, 
so this process could be suitable  to disentangle the values of $\xi$ and $\sin\theta$ once the masses of $H$ and $A$ are known, e.g., from studies of
the $e^+e^-\to HA$ cross section. 

\begin{figure}[t]
\begin{center}
\includegraphics[width=150mm]{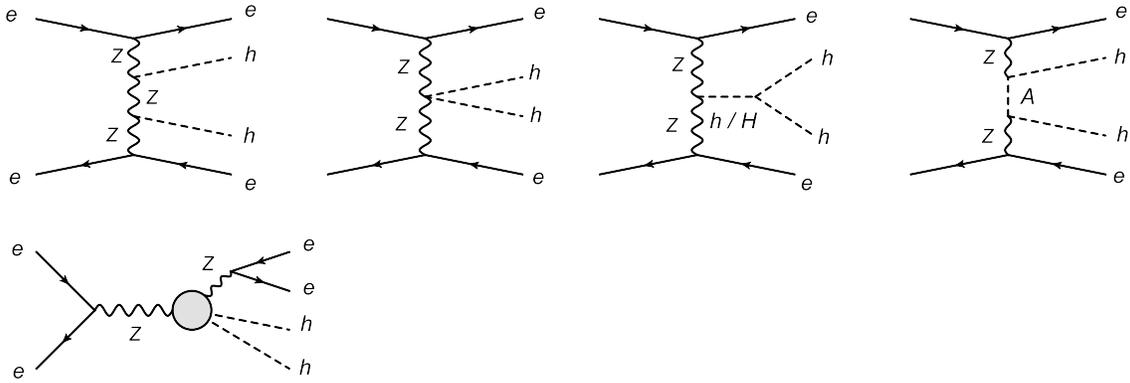}
\caption{Representative Feynman diagrams for the $e^+e^- \to e^+e^-hh$ process. The last diagram corresponds to the process $e^+e^- \to Zhh \to e^+e^-hh$ (see Fig.~\ref{diag_Zhh}). }
\label{diag_eehh}
\end{center}
\end{figure}


Next, we discuss  double-$h$ production via the VBF process. The representative Feynman diagrams are shown in Fig.~\ref{diag_eehh}. 
The HS topologies (last diagram in Fig.~\ref{diag_eehh}) play a subdominant role due to the tiny branching ratio of $Z\to e^+e^-$.
Similarly to  $Zhh$ production, the $\tan\beta$ dependence only enters via  the $\lambda_{hhh}$ and $\lambda_{Hhh}$ couplings, so that it is very small for small $\sin\theta$ values.
The most remarkable difference with respect to  $Zhh$ production is the fact that 
there is no $A$ resonance, since  only the $H$ one is possible, in the VBF process.  
Hence, it is not surprising to see the rather different dependence of the cross sections upon $\sin\theta$ and $\xi$, with respect to Fig.~\ref{cross_Zhh1}.

\begin{figure}[t]
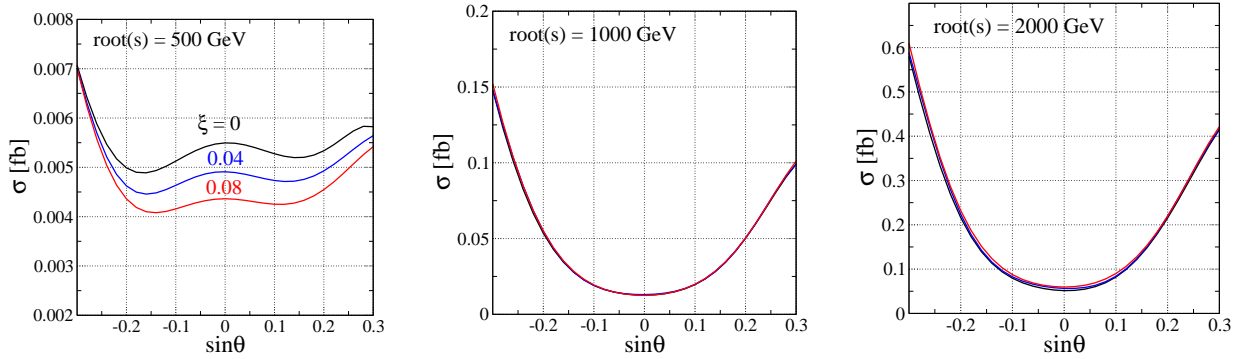

\begin{center}
\includegraphics[width=50mm]{eehh_rs500_m320_tb2.eps}\hspace{5mm}
\includegraphics[width=50mm]{eehh_rs1000_m400_tb2.eps}\hspace{5mm}
\includegraphics[width=50mm]{eehh_rs2000_m400_tb2.eps}
\caption{Cross section of the $e^+e^-\to e^+e^-hh$ process as a function of $\sin\theta$ in the C2HDM with $\tan\beta=2$ and $M =0.8 m_A^{}$. 
The collision energy and the mass ($\sqrt{s}$,$m_A^{}$) with $m_H^{}=m_A^{}$ is taken to be (500,400), (1000,500) and (2000,500) in GeV unit 
for the left, center and right panels, respectively.}
\label{cross_eehh}
\end{center}
\end{figure}

In Fig.~\ref{cross_eehh}, the  cross section for the VBF process is shown as a function of $\sin\theta$ with  $\tan\beta =2$. 
The typical behaviour is quite similar to that seen in $Zhh$ production. 
However, the difference between the C2HDM and the E2HDM evaluated for the same value of $\sin\theta$ 
is not so significant as compared to the $Zhh$ case despite the absence of the $A$ resonance.

\begin{figure}[t]
\begin{center}
\includegraphics[width=70mm]{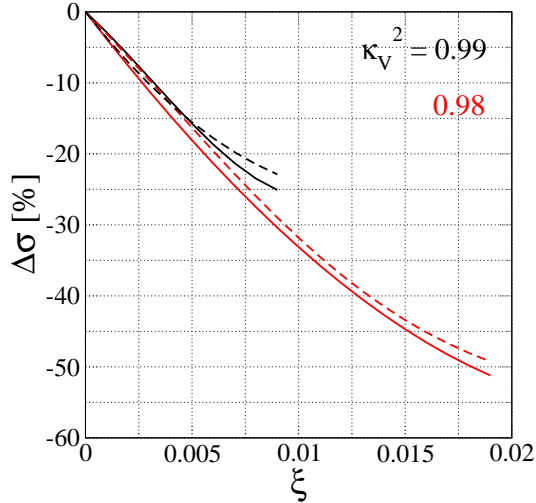}
\caption{
Deviations in the cross section $\Delta \sigma \equiv (\sigma_{\text{C2HDM}}/\sigma_{\text{E2HDM}} -1)$ for the $e^+e^- \to e^+e^-hh$ process 
at fixed $\kappa_V^2=0.99$ (black curves) and 0.98 (red curves).  
We take $\tan\beta =2$ and $\sqrt{s}=1000$ GeV. 
For each $\kappa_V^2$, we show the case of $\sin\theta>0$ (solid lines) and $\sin\theta<0$ (dashed lines). 
We only show the result allowed by the unitarity and vacuum stability bounds and by the future LHC data assuming 3000 fb$^{-1}$ with 95\% CL.  
The latter bound is for the Type-I C2HDM.}
\label{dcross_eehh}
\end{center}
\end{figure}

In Fig.~\ref{dcross_eehh}, we show the deviations in the VBF cross section as obtained in the E2HDM case relatively to the C2HDM one by taking again  
$\kappa_V^2$ = 0.99 and 0.98, in line with previous examples. 
The behaviour is similar to that seen in Fig.~\ref{dcross_Zhh}, but the 
the magnitude of the deviation can be about $-50\%$ in the case of $\kappa_V^2=0.98$ and $\sin\theta > 0$. 
Here, according to Fig.~\ref{diag_eehh}, investigation of the $hhh$, $Hhh$ and $hhZZ$ couplings 
would become possible in presence of the knowledge of the $Zhh$ one (recall that the $AhZ$ vertex is fixed by the gauge structure). 
Unlike the previously studied $Zhh$ production case though, now, 
because the cross section of the VBF process does not significantly depend upon $\tan\beta$, one is in an
excellent position to separate the values of $\xi$ and $\theta$.


\begin{figure}[t]
\begin{center}
\includegraphics[width=150mm]{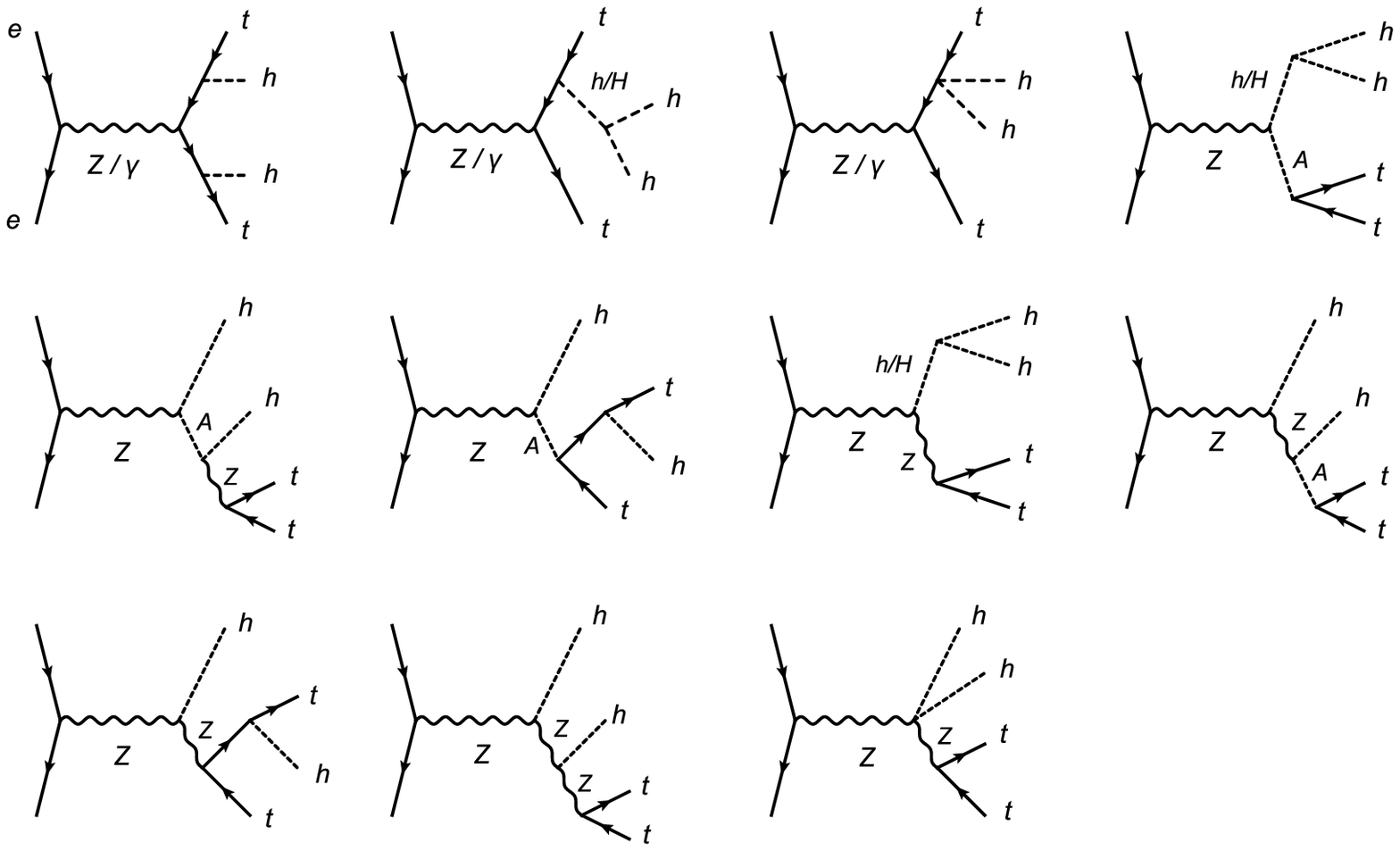}
\caption{Representative Feynman diagrams for the $e^+e^- \to t\bar{t} hh$ process. }
\label{diag_tthh}
\end{center}
\end{figure}

\begin{figure}[t]
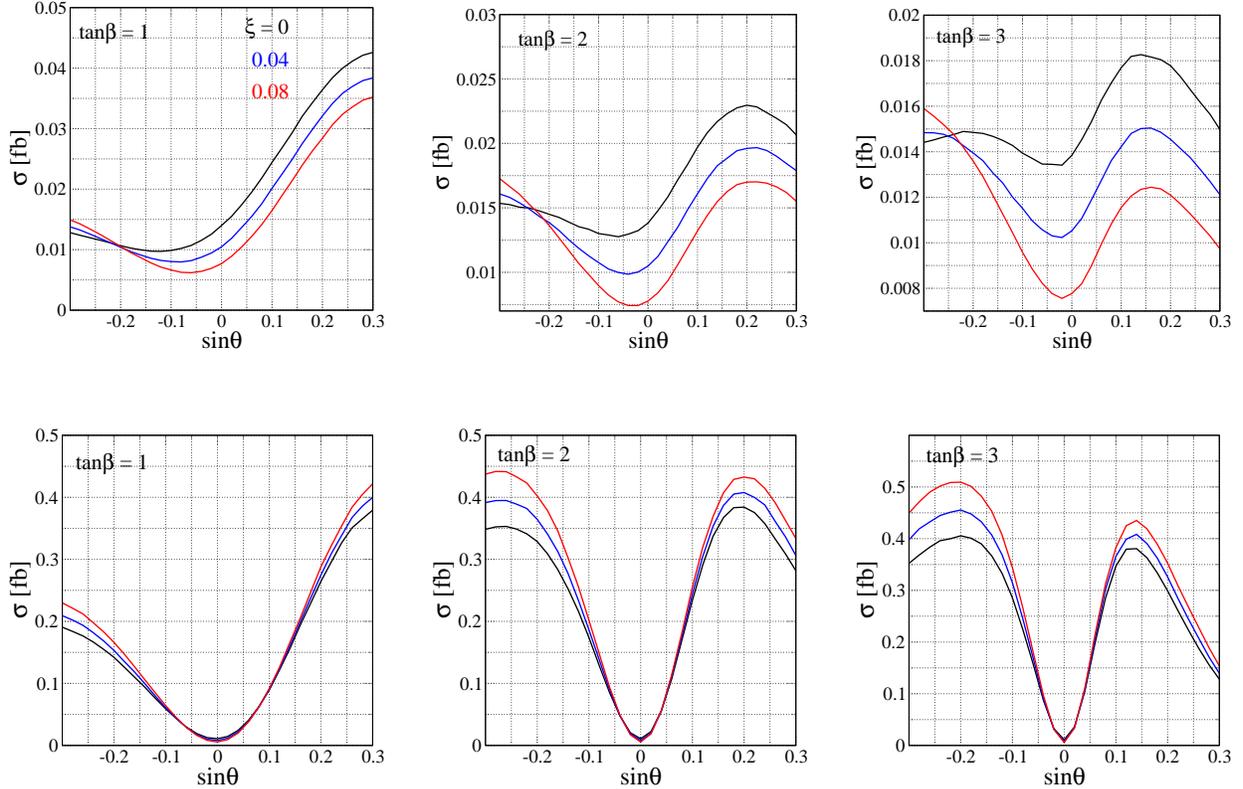

\begin{center}
\includegraphics[width=50mm]{tthh_rs1000_m400_tb1.eps}\hspace{5mm}
\includegraphics[width=50mm]{tthh_rs1000_m400_tb2.eps}\hspace{5mm}
\includegraphics[width=50mm]{tthh_rs1000_m400_tb3.eps}\\ \vspace{10mm}
\includegraphics[width=50mm]{tthh_rs2000_m400_tb1.eps}\hspace{5mm}
\includegraphics[width=50mm]{tthh_rs2000_m400_tb2.eps}\hspace{5mm}
\includegraphics[width=50mm]{tthh_rs2000_m400_tb3.eps}
\caption{Cross section of the $e^+e^-\to t\bar{t}hh$ process as a function of $\sin\theta$ in the C2HDM with $m_A^{} =m_H= 500$ GeV and $M = 0.8 m_A^{}$.
We take $\tan\beta=1$, 2 and 3 for the left, center and right panels, respectively. 
The collision energy $\sqrt{s}$ is taken to be 1000 GeV (top) and 2000 GeV (bottom). }
\label{cross_tthh}
\end{center}
\end{figure}

\begin{figure}[t]
\begin{center}
\includegraphics[width=70mm]{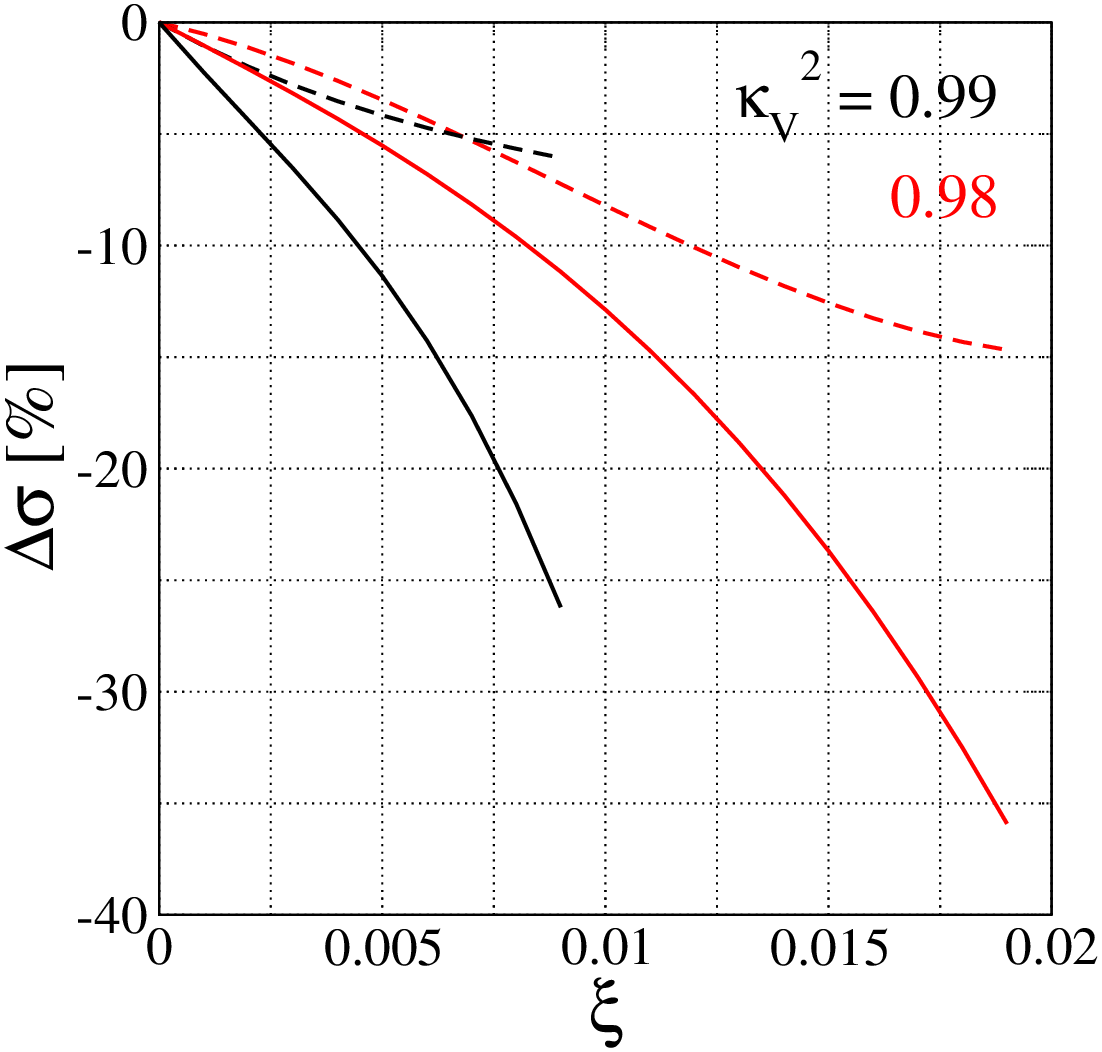}
\caption{Deviations in the cross section $\Delta \sigma \equiv (\sigma_{\text{C2HDM}}/\sigma_{\text{E2HDM}} -1)$ for the $e^+e^- \to t\bar{t}hh$ process 
at fixed $\kappa_V^2=0.99$ (black curves) and 0.98 (red curves).  
We take $\tan\beta =2$ and $\sqrt{s}=1000$ GeV. 
For each $\kappa_V^2$, we show the case of $\sin\theta>0$ (solid lines) and $\sin\theta<0$ (dashed lines). 
We only show the result allowed by the unitarity and vacuum stability bounds and by the future LHC data assuming 3000 fb$^{-1}$ with 95\% CL. 
The latter bound is for the Type-I C2HDM.}
\label{dcross_tthh}
\end{center}
\end{figure}

Finally, we investigate the $t\bar{t}hh$ production process, for which the representative Feynman diagrams are shown in Fig.~\ref{diag_tthh}. 
We note (again) that the third diagram appears  at the tree level only in  C2HDMs, not E2HDMs, as it comes from the non-linear nature of the composite Higgs interaction. 
Its effect is, however, negligibly small, because the $y_{hhtt}$ coupling given in Eq.~(\ref{Yukawas}) is proportional to $\xi$. 
Differently from the previous two double-$h$ production modes, the $\tan\beta$ dependence is important here, i.e., it enters not only $\lambda_{hhh}$ and $\lambda_{Hhh}$  
but also various Yukawa couplings such as $y_{htt}$, $y_{Htt}$ and $\tilde{y}_{Att}$ (other than indirectly via the
Higgs widths).  Therefore, this process is useful to extract independent
 information on $\tan\beta$ or to check its consistency with other processes 
if $\xi,~\sin\theta,~\tan\beta$ and the masses (plus possibly widths) 
of the extra Higgs bosons are known to some extent. 

In Fig.~\ref{cross_tthh}, we show the cross section of the $t\bar{t}hh$ process as a function of $\sin\theta$ with $m_A^{} = m_H^{} =500$ GeV and $M = 0.8 m_A^{}$. 
We take $\tan\beta=1$, 2 and 3 for the left, center and right panels, respectively. 
The cross section increases when $\sin\theta \gtrsim 0$ 
because the $H\to hh$ decay mode opens up and diagrams with the $AhZ$ vertex, e.g., the fourth topology in Fig.~\ref{diag_tthh}, become non-zero. 
Comparing the top   ($\sqrt{s}=1$ TeV) and  bottom  ($\sqrt{s}=2$ TeV) panels we see that
the cross section at $\sqrt{s}=2$ TeV is roughly one order of magnitude larger than that at $\sqrt{s}=1$ TeV  when $\sin\theta \neq 0$. 
This can be explained with the opening of  on-shell $HA$ production (again, see the fourth diagram in Fig.~\ref{diag_tthh}) with 
the subsequent  decays of $H\to hh$ and $A\to t\bar{t}$. 

Despite in both previous figures differences between the E2HDM ($\xi=0$) and C2HDM ($\xi\ne0$) cases can be large, up
to several tens of percent in the regions allowed by Tab.~\ref{sin-limit}, again, differences between the two scenarios
become very apparent if $h{VV}$ is fixed. Thus, in
 Fig.~\ref{dcross_tthh}, we show their relative cross sections of the $t\bar{t}hh$ process for a fixed value of $\kappa_V^2$=0.99 and 0.98. 
Here, we take $m_A^{} = m_H^{} = 500$ GeV, $M=0.8 m_A^{}$, $\tan\beta=2$ and $\sqrt{s}=1000$ GeV. 
As we can see, the deviation is negative and can be more than 30\% for $\kappa_V^2=0.98$ with positive $\sin\theta$. 
This is simply because of the fact that the cross section has a minimum at $\sin\theta =0$ and there are no
significant cancellations amongst the diagrams in Fig.~\ref{diag_tthh}. In contrast,  
for positive values of $\sin\theta$ the deviation can be positive, so that relative sign differences  amongst the mentioned graphs
can offset the generally negative rescaling of all vertices through $\xi$.

In short, the double-$h$ productions can be useful to access the $\xi$ and $\theta$ parameters 
by measuring the cross sections of the HS and VBF processes, since these are not simply given by the factor of $\kappa_V^2$ yet they
show little sensitivity to $\tan\beta$. 
The $t\bar{t}hh$ production is instead useful to extract $\tan\beta$ and crucial to check the self-consistency of 
either Higgs scenario, elementary or composite, given the variety of particles and interactions intervening in it. 

\section{Conclusions}

In this study, we have continued our investigation of C2HDM scenarios, initiated by Ref.~\cite{DeCurtis:2016scv}
and expanded in Ref.~\cite{DeCurtis:2016tsm} (see also \cite{DeCurtis:2016gly} for an overview), wherein the nature of all Higgs bosons is such that they are composite states, i.e.,  pNGBs from a global
symmetry breaking $SO(6)\to  SO(4)\times SO(2)$, induced explicitly by interactions between a new strong sector and the SM fields at a compositeness scale $f$. 
Furthermore, for the scalar potential, 
we assume the same general form as in the E2HDM. 
Within this construct, we have herein  proceeded to carry out a phenomenological study aiming at establishing the potential of 
 future $e^+e^-$ colliders
in disentangling the two hypotheses, E2HDM versus C2HDM. These machine environments afford one with  
a very high precision  achievable in measuring the SM-like $h$ production cross section in both  single- and double-$h$ mode, so that the rather different
   patterns  emerging in the composite scenario with respect to the elementary one may effectively be tested. We have proven this to be the case for all available modes: i.e., HS, VBF and associated production with top quarks.
Separation between the two non-minimal elementary and composite Higgs hypotheses can potentially (i.e., depending 
on the values of $\sin\theta$ and $\xi$ but irrespectively of $\tan\beta$) be achieved in all channels. In fact, for some
 combinations of  $\sin\theta$ and $\xi$, the C2HDM produces  large, and typically negative, corrections to the SM coupling strengths, up to order $-20\%$ or so, that cannot  ever be obtained in the E2HDM, thereby enabling one to promptly 
distinguish between the two scenarios. In other cases, when similar deviations can be obtained in both scenarios within expected accuracy for some (different in the two models) combination of inputs, one has to resort to a multi-dimensional 
fit assessing the CL in either hypotheses. Yet, even in this case, we expect the separation to be possible. 

We have reached these conclusions  assuming a Type-I setup in the Yukawa sector, although  we have argued that our results are independent of the
interactions of the Higgs states with fermions, as the Yukawa dependence only enters in higher orders through the width of
the heavy CP-even and CP-odd states, at least for  the $h$ production modes we have considered. Only the experimental 
constraints are in fact type dependent, yet the above prospects  about the possibility of disentangling the two 2HDM
realisations persist for all types.

This has been achieved in presence of theoretical and experimental constraints, the latter extrapolated to the end
of the LHC era, assuming both a standard (up to 300 fb$^{-1}$) and HL (up to 3000 fb$^{-1}$) setup for it. We are therefore lead to conclude that future electron-positron colliders operating between 500 and 2000 GeV, of which there exist several prototypes (such as  the aforementioned ILC 
\cite{Phinney:2007zz,Djouadi:2007ik,Phinney:2007gp,Behnke:2007gj,Brau:2007zza,BrauJames:2007aa,Behnke:2013lya,Adolphsen:2013kya,Adolphsen:2013jya,Behnke:2013xla,Baer:2013cma}, but also the 
 Compact Linear Collider (CLiC)~\cite{Aicheler:2012bya} and  Future Circular Collider $e^+e^-$ (FCC-ee)
\cite{Gomez-Ceballos:2013zzn}), 
are the ideal testing ground to confirm or disprove the existence in Nature of
either a E2HDM or C2HDM as the underlying dynamics of electro-weak symmetry breaking.

\noindent
\section*{Acknowledgments}
\noindent 
The work of SM was financed in part through the NExT Institute and by the STFC Consolidated Grant ST/J000391/1.
EY was supported by the Ministry of National Education of Turkey.

\newpage

\begin{appendix}

\section*{Appendix}

\begin{table}[h!]
\begin{center}
\begin{tabular}{ll}
\begin{tabular}{|c|c|c|}\hline
Channel &  Refs. \\ \hline

\makecell{$\tau_{\text{had}} \tau_{\text{had}}$ (VBF, 8 TeV)\\$\tau_{\text{lep}} \tau_{\text{had}}$ (boosted, 8 TeV)\\ $\tau_{\text{lep}} \tau_{\text{had}}$ (VBF, 8 TeV)\\$\tau_{\text{lep}} \tau_{\text{lep}}$ (boosted, 8 TeV)\\ $\tau_{\text{lep}} \tau_{\text{lep}}$ (VBF, 8 TeV)} & \cite{tau} \\ \hline
WW (VBF enhanced, 8TeV) & \cite{ww} \\ \hline
ZZ (VBF, 8TeV) & \cite{zz} \\ \hline
\makecell{multilepton\\1$\ell 2 \tau_{\text{had}}$ ($t\bar{t}H$, 8 TeV)\\ 2$\ell 0 \tau_{\text{had}}$ ($t\bar{t}H$, 8 TeV)\\2$\ell 1\tau_{\text{had}}$ ($t\bar{t}H$, 8 TeV)\\3$\ell$ ($t\bar{t}H$, 8 TeV)} & \cite{multilep} \\ \hline

$ b\bar{b} $ (Vh, 8 TeV) with 1-lepton and 2-lepton channels & \cite{bb} \\ \hline
$ b\bar{b} $ (Vh, 13 TeV) with 0-lepton,  1-lepton and 2-lepton channels & \cite{13bb} \\ \hline
\end{tabular}
\end{tabular}
\vspace{5mm}
\caption{List of the Higgs data samples used in the $ \Delta\chi^{2} $ calculations in  Figs.~\ref{HB/HS}--\ref{HB/HS2}.}
\label{channels}
\end{center}
\end{table}

\begin{table}[h]
\resizebox{\columnwidth}{!}{
\scalebox{0.12}{
\begin{tabular}{ll}
\begin{tabular}{|c|c|c|c|c|c|}\hline
Model Type &  $ \tan\beta=1 $ &$ \tan\beta=2 $& $ \tan\beta=3 $\\ \hline
Type I & \makecell{$ pp \to  H \to  ZZ \to  4\ell $ \cite{atlas1}\\
$qQ \to q'Q'h \to WW \to 2\ell 2\nu$  \cite{cms3}}& \makecell{$ pp \to  H \to  ZZ \to  4\ell $ \cite{atlas1} \\
$qQ \to q'Q'h \to WW \to 2\ell 2\nu$  \cite{cms3}}& \makecell{$ pp \to  H \to  ZZ \to  4\ell $ \cite{atlas1}\\
$qQ \to q'Q'h \to WW \to 2\ell 2\nu$  \cite{cms3}} \\ \hline

Type II    &  \makecell{$ pp \to  H \to  ZZ \to  4\ell $ \cite{atlas1}\\$qQ \to q'Q'h \to WW \to 2\ell 2\nu$  \cite{cms3}} & \makecell{$ pp \to  H \to  ZZ \to  4\ell $ \cite{atlas1}\\ $ pp \to  h \to  WW^{\ast} \to  \ell\nu\ell\nu $ \cite{atlas2}\\$qQ \to q'Q'h \to WW \to 2\ell 2\nu$  \cite{cms3}\\$ pp \to  h \to  ZZ \to  4\ell $ \cite{cms4}} &\makecell{$ pp \to  H \to  ZZ \to  4\ell $ \cite{atlas1}\\
$qQ \to q'Q'h \to WW \to 2\ell 2\nu$  \cite{cms3}\\$ pp \to  h \to  ZZ \to  4\ell $ \cite{cms4}} \\ \hline

Type X   & \makecell{$ pp \to  H \to  ZZ \to  4\ell $ \cite{atlas1}\\$qQ \to q'Q'h \to WW \to 2\ell 2\nu$ \cite{cms3}}&\makecell{ $ pp \to  H \to  ZZ \to  4\ell $ \cite{atlas1}\\$qQ \to q'Q'h \to WW \to 2\ell 2\nu$ \cite{cms3}}&\makecell{$ pp \to  H \to  ZZ \to  4\ell $\cite{atlas1}\\$qQ \to q'Q'h \to WW \to 2\ell 2\nu$\cite{cms3}}\\  \hline

Type Y &\makecell{ $ pp \to  H \to  ZZ \to  4\ell $ \cite{atlas1}\\$qQ \to q'Q'h \to WW \to 2\ell 2\nu$ \cite{cms3}\\ gg $ \to  \phi (h,H) \to  \tau \tau $  \cite{cms5}}&\makecell{$ pp \to  H \to  ZZ \to  4\ell $ \cite{atlas1}\\ $ pp \to  H \to  hh \to  4b $ \cite{cms1}\\$ pp \to  H \to  WW^{\ast} \to  \ell\nu\ell\nu $ \cite{atlas2}\\$ pp \to  h \to  ZZ \to  4\ell $ \cite{cms4}}& \makecell{  $ pp \to  H \to  hh \to  4b $ \cite{cms1}\\$qQ \to q'Q'h \to WW \to 2\ell 2\nu$ \cite{cms3}\\$ pp \to  h \to  ZZ \to  4\ell $ \cite{cms4}}\\ \hline
\end{tabular}
\end{tabular}
}}
\vspace{5mm}
\caption{Higgs search channels most responsible for excluding parameter regions  in Figs.~\ref{HB/HS}--\ref{HB/HS2}. }
\label{exc-channels}
\end{table}

\end{appendix}

\end{document}